\documentclass[journal]{IEEEtran}   
\usepackage{multirow}
\usepackage{diagbox}
\usepackage{amssymb}
\usepackage{amsmath}
\usepackage{graphicx}
\usepackage{cite}
\usepackage{citesort}
\usepackage{subfigure}
\usepackage{graphicx,epstopdf}
\usepackage{epsfig}	
\usepackage{cite,graphicx,amsmath,amssymb}
\usepackage{comment}
\usepackage{booktabs}
\usepackage{amssymb}
\usepackage{amsmath}
\usepackage{cite}
\usepackage{url}
\usepackage{xcolor}
\usepackage{cite,graphicx,amsmath,amssymb}
\usepackage{subfigure}
\usepackage{citesort}
\usepackage{fancyhdr}
\usepackage{mdwmath}
\usepackage{mdwtab}
\usepackage{caption}
\usepackage{amsthm}
\usepackage{lipsum}
\usepackage{algorithm}
\usepackage{algorithmic}
\usepackage{multirow}
\allowdisplaybreaks[4]

\newtheorem{theorem}{Theorem}

\newtheorem{lemma}{Lemma}

\newtheorem{corollary}{Corollary}

\newtheorem{remark}{Remark}  

\makeatletter
\def\ScaleIfNeeded{%
\ifdim\Gin@nat@width>\linewidth \linewidth \else \Gin@nat@width
\fi } \makeatother

\makeatletter

\renewcommand{\maketag@@@}[1]{\hbox{\m@th\normalsize\normalfont#1}}%

\makeatother

\begin{document}

\title{\Huge{Federated Learning in Active STARS-Aided Uplink Networks}}

\author{Xinwei~Yue,~\IEEEmembership{Senior Member,~IEEE,} Xinning~Guo,~\IEEEmembership{Graduate Student Member,~IEEE,} Xidong~Mu,~\IEEEmembership{Member,~IEEE,} Jingjing~Zhao,~\IEEEmembership{Member,~IEEE,}  Peng~Yang,~\IEEEmembership{Member,~IEEE,} Junsheng Mu, and  Zhiping~Lu
\thanks{X. Yue and X. Guo are with the Key Laboratory of Information and Communication Systems, Ministry of Information Industry and also with the Key Laboratory of Modern Measurement $\&$ Control Technology, Ministry of Education, Beijing Information Science and Technology University, Beijing 102206, China (e-mail: \{xinwei.yue, xinning.guo\}@bistu.edu.cn).}
\thanks{X. Mu is with the Centre for Wireless Innovation (CWI), Queen's University Belfast, Belfast, BT3 9DT, U.K. (e-mail: x.mu@qub.ac.uk).}
\thanks{J. Zhao and P. Yang are with the School of Electronic and Information Engineering, and also with State Key Laboratory of CNS/ATM, Beihang University, Beijing 100191, China (e-mail: \{jingjingzhao, yangp09\}@buaa.edu.cn).}
\thanks{J. Mu is with the School of Informationand Communication Engineering, Beijing University of Posts andTelecommunications, Beijing 100876, China (e-mail: mujs@bupt.edu.cn).}
\thanks{Z. Lu is with State Key Laboratory of Wireless Mobile Communications, China Academy of Telecommunications Technology (CATT), Beijing 100191, China  (e-mail:  luzp@cict.com).}
}

\maketitle
\begin{abstract}
Active simultaneously transmitting and reflecting surfaces (ASTARS) have attracted growing research interest due to its ability to alleviate multiplicative fading and reshape the electromagnetic environment across the entire space.
In this paper, we utilise ASTARS to assist the federated learning (FL) uplink model transfer and further reduce the number of uploaded parameter counts through over-the-air (OTA) computing techniques.
The impact of model aggregation errors on ASTARS-aided FL uplink networks is characterized.
We derive an upper bound on the aggregation error of the OTA-FL model and quantify the training loss due to communication errors.
Then, we define the performance of OTA-FL as a joint optimization problem that encompasses both the assignment of received beams and the phase shifting of ASTARS, aiming to achieve the maximum learning efficiency and high-quality signal transmission.
Numerical results demonstrate that: i) The FL accuracy in ASTARS uplink networks are enhanced compared to that in state-of-the-art networks; ii) The ASTARS enabled FL system achieves the better learning accuracy using fewer active units than other baseline, especially when the dataset is more discrete; and iii) FL accuracy improves with higher amplification power, but excessive amplification makes thermal noise the dominant source of error.
\end{abstract}
\section{Introduction}
As wireless communication technology rapidly evolves, sixth-generation (6G) wireless networks face considerable challenges related to extensive connectivity, high-speed data transmission, and low-latency service delivery \cite{zhang20196g,giordani2020toward}. To realize the network vision of ubiquitous and intelligent connectivity, 6G networks are capable of providing the offer services for various scenarios such as artificial intelligence, smart cities, and smart industries \cite{you2021towards,10918743}. One major challenge is to support massive data transfer and private data transfer services with constrained communication resources to meet the smart service requirements of 6G networks \cite{park2019,saad2020}.

With the rapid development of machine learning and artificial intelligence technologies, there is an increasing demand for natural language processing \cite{li2018deep} and intelligent image processing \cite{razzak2018deep} on mobile devices. Traditional machine learning approaches often require centralized data aggregation for model training \cite{jordan2015machine}, which incurs high communication overhead and raises potential data privacy concerns \cite{zhu2020toward}. In contrast, federated learning (FL) mitigates these issues by transmitting only gradient updates to the central server, significantly reducing communication overhead while preserving local data privacy \cite{mammen2021federated}. This renders FL especially appropriate for implementation on Internet of Things devices and mobile platforms in contexts like smart cities and intelligent industries \cite{yang2019}.
The accuracy of FL models can be further enhanced through model structuring techniques \cite{konevcny2016federated}, which have become widely adopted in various fields. By selectively uploading gradient information, the authors of \cite{icdcs2019} demonstrate that optimizing communication resources significantly improves the performance of FL. The authors of \cite{Luo2021} proposed an adaptive FL algorithm, which selectively uploads gradients from heterogeneous devices, reducing communication and computational overhead in mobile edge computing scenarios and enhancing overall system efficiency. In \cite{amiri2020}, the authors introduced a distributed stochastic gradient descent approach tailored for wireless fading channels, enhancing both the learning performance and resilience of FL in such environments. In \cite{FM2024}, the authors conducted an in-depth study on pipeline parallelism, data parallelism, and multi-modal learning, making a significant contribution to the development of foundational models for distributed learning.

With the popularization of wireless networks and the surge in the number of IoT devices, the application of FL in wireless communication environments has received widespread attention \cite{samarakoon2019distributed,niknam2020federated}. The challenges faced by FL in wireless communication environments mainly include bandwidth limitations, channel instability, delay sensitivity, and device heterogeneity \cite{wen2023survey}. In order to address these challenges, researchers have proposed a variety of methods to optimize the performance of FL in wireless environments \cite{FL2021TONG}. For example, by adopting a channel-aware client selection strategy, the research of has shown that it can effectively reduce communication overhead and improve system efficiency \cite{Chen2021}. In addition, for wireless fading channels, the distributed stochastic gradient descent method proposed by significantly improves the performance of FL in unstable wireless channels \cite{kassab2022federated}. In \cite{10742117}, the authors proposed a time synchronization and channel estimation method based on orthogonal space-time, which is believed to improve model aggregation performance. With the introduction of wireless computing and non-orthogonal multiple access (NOMA) technology, the communication delay in FL has been further reduced, and the efficiency of FL has been improved by utilizing the transmission of multiple signals within the same resource block \cite{NI2022FL}.
 To further reduce uplink communication delays, the authors of \cite{yang2020federated} proposed an over-the-air (OTA) computation function within wireless FL, aiming to enhance efficiency. By utilizing NOMA, OTA computing transmits signals over the same physical resources, facilitating FL gradient computation via signal superposition \cite{AIRCOM2021}. This method demonstrated that increasing the number of users utilizing the same physical resource block does not result in greater communication delays \cite{AIRC2020}. However, OTA computing encounters challenges related to discrete problems in \cite{AIRC2023}, which can significantly affect accuracy of FL processes over wireless communication networks. To support OTA computing, the essential power control mechanisms allowed users with poorer conditions to transmit the signals at higher power in \cite{cao2020optimized}, which leads to model aggregation errors. However, this approach can result in larger model aggregation errors from users with poor channel conditions.

Reconfigurable intelligent surfaces (RIS) technology stands as a cornerstone in the advancements toward 6G, promising substantial improvements in system capacity, coverage expansion, and signal strength enhancement\cite{di2020radio,pris2020,LIU2021RIS}. Traditionally, passive RIS (PRIS) mainly comprised passive units tasked with manipulating signal phase and amplitude to mold the electromagnetic environment \cite{basar2019wireless}. This reduction in transmit power requirement not only lowers the energy consumption of the system but also mitigates electromagnetic interference to the surrounding environment \cite{greenris}.  Since RIS can mitigate unfavorable channel conditions, recent studies have shown that RIS can significantly reduce errors in OTA computing tasks \cite{RCRIS2021}. Since PRIS is designed to operate solely in reflection mode, it inevitably fails to address certain use scenarios. Passive simultaneously transmitting and reflecting surfaces (PSTARS) utilize multiple elements to reflect and transmit signals simultaneously, providing seamless services to users throughout a space \cite{kha2023STARRIS}. The authors of \cite{wang2023} model the energy consumption of PSTARS-assisted terahertz communication systems and demonstrate that PSTARS offers superior performance improvements over PRIS in enhancing communication efficiency.
However, PRIS/PSTARS rely on passive elements, which exacerbates channel cascade fading, particularly in base station (BS)-PSTARS-users channels, where multiplicative fading is more severe than in directly connected channels \cite{zhang2022active}.  The presence of cascading channel fading further increases the bit error rate, ultimately affecting the reliability of data transmission \cite{li2021CHANEL}.

To tackle these challenges, active RIS (ARIS) is capable of significantly enhancing signal transmission while effectively mitigating fading issues that often arise in wireless communication networks \cite{Long2021ARIS}. ARIS, featuring a comprehensive array of active elements, is engineered to address the multiplicative fading that is characteristic of channel cascades \cite{ARIS2021}. The ARIS-assisted wireless communication system surpasses the PRIS system in spectral efficiency and coverage when the system budget is established \cite{POWERARIS2022}. The authors of \cite{ARISNOMA2024} examined the outage probability and ergodic rate, finding that ARIS-NOMA networks outperform ARIS-orthogonal multiple access networks. In \cite{DAS2023109581}, the authors illustrate that machine learning can enhance the communication performance of wireless networks by dynamically adjusting the parameters of ARIS.
In order to compensate for ARIS's inability to serve the full space regret, ASTARS has received a lot of attention from academia and industry \cite{ASTARS2023}. ASTARS is capable of transmitting signals in various directions through a combination of reflection and transmission, and it also employs active units to amplify these signals, thereby overcoming multiplicative fading \cite{rate2023}. Compared to PSTARS, ASTARS offer several advantages: It provides larger coverage areas, mitigate transmission errors by overcoming multiplicative interference of cascaded channels, and exhibit higher energy efficiency. A seminal work proposed a hardware model for ASTARS and designed an ASTARS-assisted downlink communication system \cite{ASTARS2023}. Subsequent studies demonstrated that ASTARS-assisted systems achieve higher throughput and increased capacity utilization \cite{rate2023}. Moreover, adjusting the phase shift of signals in ASTARS-NOMA downlink networks improves outage probability and ergodic data rate \cite{yueandxie2024}. By addressing the limitations of passive systems and leveraging active elements, ASTARS present a promising solution for enhancing modern communication networks.

\begin{table*}[t]
\centering
\caption{Our contributions in contrast to the state-of-the-art.}
\resizebox{\textwidth}{!}{%
\begin{tabular}{|l|c|c|c|c|c|c|}
\hline
 & \cite{NI2022FL} & \cite{AIRCOM2021} &\cite{RCRIS2021} &\cite{DAS2023109581} & \cite{yueandxie2024} & Proposed \\
\hline
Full-space signal transmission  & \checkmark & $\times$ & $\times$ & $\times$ & \checkmark & \checkmark \\
\hline
Signal amplification units  & $\times$ & \checkmark & $\times$& \checkmark & \checkmark & \checkmark \\
\hline
Consideration of amplification factor impact & $\times$ & $\times$ & $\times$ & $\times$ & \checkmark & \checkmark \\
\hline
Support for FL  & \checkmark & \checkmark, but not mentioned & \checkmark & \checkmark, but not mentioned & $\times$ & \checkmark \\
\hline
Use of OTA computation  & \checkmark & \checkmark & \checkmark & $\times$ & $\times$ & \checkmark \\
\hline
Quantification of model aggregation error  & \checkmark & $\times$ & $\times$ & $\times$ & $\times$ & \checkmark \\
\hline
\end{tabular}%
}
\label{tab:comparison}
\end{table*}
\subsection{Motivations and Contributions}
While the above-mentioned research provides a solid foundation for comprehending PSTARS and FL techniques, ASTARS-assisted FL remains an emerging area of study. ASTARS, which leverages advanced phased-array technology for signal amplification, holds significant potential in improving the efficiency of wireless communication systems, particularly in the context of FL. To the best of our knowledge, the integration of ASTARS-assisted FL, specifically in the uplink OTA computation system, has not been comprehensively explored in prior works. While prior research has addressed the enhancement of signal quality through various signal processing techniques, the unique advantages offered by ASTARS, such as mitigating multiplicative fading and enabling the simultaneous transmission of multiple signals OTA, remain underutilized in FL scenarios.
Motivated by this gap in existing literature, this paper seeks to investigate the impact of ASTARS on FL accuracy within the uplink network framework. The integration of ASTARS enables simultaneous transmission of signals from a large set of users to the BS, which facilitates the efficient and scalable operation of FL in wireless networks. By leveraging ASTARS's ability to amplify incident signals and mitigate fading, we also address the discretization challenges typically encountered in OTA computation, which have been a major hurdle in previous works on wireless FL systems. The main contributions of this paper can be summarized below, which are explicitly contrasted to the relevant state-of-the-art in Table \ref{tab:comparison}. 
\begin{itemize} 
\item We introduce a novel ASTARS-assisted uplink network framework for FL, leveraging the OTA computation technique to enable simultaneous signal transmission from multiple users to the BS. We derive a closed-form expression for model aggregation errors in this context and propose an efficient method for mitigating these errors in large-scale FL systems. 
\item We establish an upper bound on the learning loss by accounting for model aggregation errors in the ASTARS-assisted FL uplink network. Through a detailed analysis, we find that the primary cause of slower convergence rates is the error introduced during the aggregation of locally trained models at the BS. We provide a quantitative framework for analyzing these losses and offer insights into their mitigation. 
\item We investigate the learning rate optimization problem under ASTARS-assisted FL, emphasizing the critical role of ASTARS phase shifts. A novel algorithm based on successive convex approximation (SCA) is proposed to jointly optimize the ASTARS phase shifts and BS beamforming, significantly reducing aggregation errors and improving convergence performance.
\item We demonstrate that the proposed ASTARS-assisted system outperforms state-of-the-art baselines in learning accuracy, achieves better performance with fewer active units on heterogeneous data, and reveals the trade-off between amplification power and thermal noise impact. 
\end{itemize}

\subsection{Organization and Notation}
The structure of the paper is organized as follows. Section II outlines the system model of ASTARS uplink networks with model aggregation for FL. Section III examines the performance of FL and formulates the optimization problem. In Section IV, the SCA algorithm is applied to jointly optimize the ASTARS phase shift and received beamforming. Section V presents the simulation results and numerical analysis, while Section VI concludes the paper. A collection of mathematical proofs is included in the Appendix.

\textit{Notations}: 
Scalars, vectors, and matrices are denoted by lower-case, bold-face lower-case, and bold-face upper-case letters, respectively. 
${\mathbb{C}^{J \times Q}}$ denotes the spaece of $J \times Q$ complexvalued vectors.
${f_X}\left(  \cdot  \right)$ and ${F_X}\left(  \cdot  \right)$ represent the probability density function and cumulative distribution function of a random variable, respectively.
${\Bbb E}\left(  \cdot  \right)$ and ${\Bbb D}\left(  \cdot  \right)$ indicate the operations for expectation and variance, respectively. Additionally, ${\left(  \cdot  \right)^T}$ and ${\left(  \cdot  \right)^H}$ refer to the transpose and conjugate transpose operations, respectively.

\begin{figure}[!h]
    \begin{center}
        \includegraphics[width=3.4in]{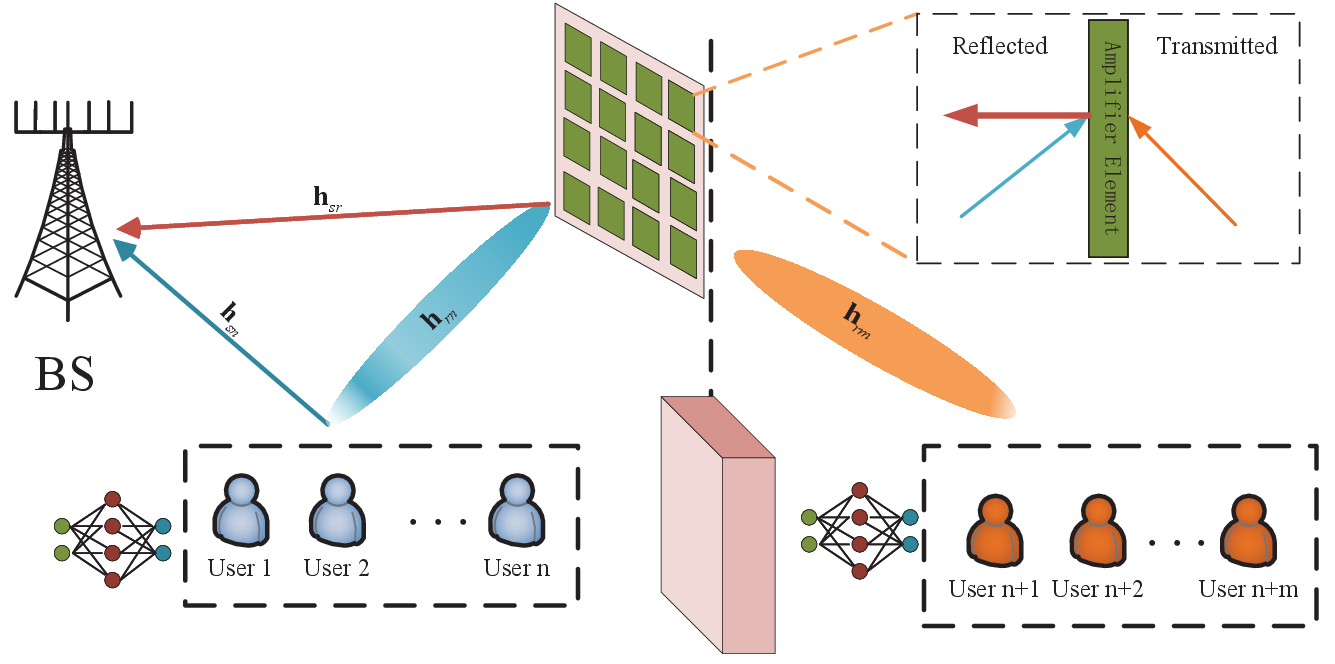}
        \caption{System model of ASTARS-assisted FL networks.}
        \label{Fig. 1}
    \end{center}
\end{figure}
\section{System Model}\label{System Model}
We examine an ASTARS-assisted FL network, as depicted in Fig. 1, where the signals from users are refracted and amplified before reaching the BS.
The BS has $J$ antennas, while each user utilizes a single-antenna device. The ASTARS consists of $Q$ active refracted and amplified elements. The distributed users are FL users. The ASTARS divides the signal propagation space into reflection space and transmitting space. More precisely, the set of FL users situated in the reflection space is represented as ${U_n} = \left\{ {{U_1}, \cdots ,{U_n}, \cdots ,{U_N}} \right\}$ while the users located in the transmitting space are denoted by ${U_m} = \left\{ {{U_1}, \cdots ,{U_m}, \cdots ,{U_M}} \right\}$,  with the condition that ${U_m} \cap {U_n} = \emptyset $. In practice, the sets of users $N$ and $M$ can be readily determined, provided the number of users is known and their locations are fixed. For ease of presentation, we let $\Phi  = N \cup M$.
Unlike traditional reflection RIS, ASTARS enables omni-directional radiation by incorporating equivalent electric and magnetic currents into the hardware design of each element. Additionally, ASTARS elements are equipped with active amplifiers to boost the incident signal, counteracting the attenuation caused by multiplicative fading. The complex channel coefficients from the BS to ${U_n}$, the BS to ASTARS, and then from ASTARS to ${U_\varphi }, {\varphi  \in \Phi }$ are denoted by ${{\mathbf{h}}_{sn}} \in {\mathbb{C}^{J \times 1}}$, ${{\mathbf{h}}_{sr}} \in {\mathbb{C}^{J \times Q}}$, and ${{\mathbf{h}}_{r\varphi}} \in {\mathbb{C}^{Q \times 1}}$, ${\varphi  \in \Phi }$. For practical scenarios, ASTARS uplink communication links experience Rician fading. The terms ${\mathbf{h}}_{sr}{{\mathbf{\Theta }}^n}{{\mathbf{h}}_{rn}}$ and ${\mathbf{h}}_{sr}{{\mathbf{\Theta }}^m}{{\mathbf{h}}_{rm}}$ represent the cascaded complex channel coefficients from the BS to ASTARS, and then to ${U_n}$ and ${U_m}$, respectively.  ${{\mathbf{\Theta }}^n} = \sqrt {\lambda {\beta _n}} {\text{diag}}\left( {{e^{j\theta _1^n}}, \cdots ,{e^{j\theta _q^n}}, \cdots ,{e^{j\theta _Q^n}}} \right)$ and ${{\mathbf{\Theta }}^m} = \sqrt {\lambda {\beta _m}} {\text{diag}}\left( {{e^{j\theta _1^m}}, \cdots ,{e^{j\theta _q^m}}, \cdots ,{e^{j\theta _Q^m}}} \right)$ denote the reflection and transmission phase-shifting amplification matrices of ASTARS, respectively. To simplify the analysis, it is assumed that all ASTARS elements share the same amplification factor $\lambda $. The reflection and transmission amplitude coefficients are denoted as ${\beta _r}$ and ${\beta _t}$, with the condition ${\beta _n} + {\beta _m} \leqslant 1$. The phase shifts of the reflection and transmission responses for the $q$-th element are represented by $\theta _q^n,\theta _q^m \in \left[ {\left. {0,2\pi } \right)} \right.$, and they are independent of each other. Perfect CSI is required for users to perform coherent demodulation.
\subsection{Signal Model of FL Users}
In ASTARS uplink networks, distributed devices often limit their transmit power to conserve energy, leading to weaker signals received at the BS. More critically, signals from certain devices may be treated as noise, resulting in data loss errors. The phase shifts of ASTARS are kept constant throughout the FL model aggregation process. Reflection and transmission users utilize OTA computation to accomplish the aggregation of the FL model. Once the user signals reach ASTARS, the OTA computation technique ensures that the combined reflection and transmission signals are delivered to the BS simultaneously. At this point, the received signal expressions at the BS can be formulated as
\begin{align}\label{BS receive signal}
{{\bf{y}}_{{\text{BS}},t}} = &\sum\nolimits_{n = 1}^N {{{\mathbf{h}}_{sn}}\sqrt {{p_n}} {x_n}}  + {{\mathbf{h}}_{sr}}{{\mathbf{\Theta }}^n}\sum\nolimits_{n = 1}^N {{{\mathbf{h}}_{rn}}\sqrt {{p_n}} {x_n}} \notag\\
 &+ {{\mathbf{h}}_{sr}}{{\mathbf{\Theta }}^m}\sum\nolimits_{m = 1}^M {{{\mathbf{h}}_{rm}}\sqrt {{p_m}} {x_m}} + {{\mathbf{h}}_{sr}}{{\mathbf{\Theta }}^n}{{\mathbf{n}}_s}\notag \\
 &+ {{\mathbf{h}}_{sr}}{{\mathbf{\Theta }}^m}{{\mathbf{n}}_s} + {{{\mathbf{\tilde n}}}_0},
\end{align}
where $p_n$ and $p_m$ denote $U_n$'s and $U_m$'s transmit power, respectively, $x_n$ and $x_m$ denote the signals of $U_n$ and $U_m$, respectively. ${{\bf{n}}_s} = {\left[ {n_s^1, \cdots ,n_s^q, \cdots ,n_s^Q} \right]^H}$ represents the noise matrix generated by the ASTARS elements, where each $n_s^q$ follows $ {\cal C}{\cal N}\left( {0,\sigma _S^2} \right)$. ${{\bf{n}}_0} = {\left[ {n_0^1, \cdots ,n_0^j, \cdots ,n_0^J} \right]^H}$ stands for white Gaussian noise and $n_0^j \sim {\cal C}{\cal N}\left( {0,\sigma _0^2} \right)$. The channel coefficients from the BS to ASTARS, and subsequently from ASTARS to ${U_\varphi }$, are expressed as
${\bf{h}}_{sr} = \sqrt {{\eta _0}d_{sr}^{ - \alpha }} {\left[ {{\bf{h}}_{sr}^1, \cdots ,{\bf{h}}_{sr}^j, \cdots ,{\bf{h}}_{sr}^J} \right]^H}$ with ${\bf{h}}_{sr}^j = \sqrt {{\eta _0}d_{sr}^{ - \alpha }} \left[ {h_{sr}^1, \cdots ,h_{sr}^q, \cdots ,h_{sr}^Q} \right]$ and ${\mathbf{h}}_{r\varphi } = \sqrt {{\eta _0}d_{r\varphi }^{ - \alpha }} {\left[ {h_{r\varphi }^1, \cdots ,h_{r\varphi }^q, \cdots ,h_{r\varphi }^Q} \right]^H}$, respectively, where $h_{sr}^q = \sqrt {\frac{\kappa }{{\kappa  + 1}}}  + \sqrt {\frac{1}{{\kappa  + 1}}} \tilde h_{sr}^q$ with $\tilde h_{sr}^q \sim {\cal C}{\cal N}\left( {0,1} \right)$ and $h_{r\varphi }^q = \sqrt {\frac{\kappa }{{\kappa  + 1}}}  + \sqrt {\frac{1}{{\kappa  + 1}}} \tilde h_{r\varphi }^q$ with $\tilde h_{r\varphi }^q \sim {\cal C}{\cal N}\left( {0,1} \right)$. The parameter $\kappa$ represents the Rician factor, while $\alpha $ denotes the path loss exponent. The term ${\eta _0}$ indicates the path loss, ${d_{sr}}$ represents the distance from the BS to ASTARS, and ${d_{r\varphi }}$ corresponds to the distance from ASTARS to ${U_\varphi }$.

Assuming no loss of generality, we consider the previously mentioned signals to be statistically independent. They fulfill the conditions $\mathbb{E}\left[ {{{\left| {{x_\varphi }} \right|}^2}} \right] = 1$, $\varphi  \in \Phi$ and $\mathbb{E}\left[ {{x_u}{x_v}} \right] = 0$, $\forall u \ne v$. In the $t$-th round of training, the transmit power budget of the distributed devices should all be satisfied:
$\mathbb{E}\left[ {{{\left| {{p_\varphi }{x_\varphi }} \right|}^2}} \right] = {\left| {{p_\varphi }} \right|^2} \leqslant {P_\varphi },\varphi  \in \Phi $, where $P_\varphi$ stands the highest transmit power of the ${U_\varphi }$. 

It is important to highlight that the signal transmitted by the distributed device will be amplified by the ASTARS. As a result, the power of the signal received at the BS differs from the power of the signal transmitted by ${U_\varphi }$. During the $t$-th round of training, round of training, the average power of all devices must be satisfied
\begin{align}\label{average transmit power}
\frac{1}{{N + M}}\sum\nolimits_{\varphi  = 1}^\Phi  \mathbb{E} \left[ {{{\left| {{p_\varphi }{x_\varphi }} \right|}^2}} \right] = \frac{1}{{N + M}}\sum\nolimits_{\varphi  = 1}^\Phi  {{{\left| {{p_\varphi }} \right|}^2}}  \leqslant {\bar P_\varphi }.
\end{align}
\subsection{FL Framework}
The goal of learning is to minimize an empirical loss function, which can be expressed as
\begin{align}\label{learning objevtive}
\mathop {\min }\limits_{{\mathbf{w}} \in {\mathbb{R}^{D \times 1}}} F\left( {\mathbf{w}} \right) = \frac{1}{K}\sum\nolimits_{k = 1}^K {f\left( {{\mathbf{w}};{{\mathbf{x}}_k},{y_k}} \right)},
\end{align}
where ${\mathbf{w}}$ is defined as the $D$-dimensional model parameter vector; and $K$ denotes the total number of training samples. When we input feature vector x to $F\left( {\mathbf{w}} \right)$,  $y$ will be output. Assuming that the training samples are distributed among edge devices, ${K_\varphi }$ training samples on the $\varphi $-th device and $\sum\nolimits_{\varphi  = 1}^\Phi  {{K_\varphi } = K} ,{\varphi  \in \Phi }$. The loss function for a sample pair $\left( {{{\mathbf{x}}_k},{y_k}} \right)$ is denoted as $f\left( {{\mathbf{w}};{{\mathbf{x}}_k},{y_k}} \right)$. We use ${\mathcal{D}_\varphi } = \left\{ {\left( {{{\mathbf{x}}_{\varphi k}},{y_{\varphi k}}} \right):1 \leqslant k \leqslant {K_\varphi }} \right\}$ to describe local data set for the $\varphi $-th user, and $\left| {{\mathcal{D}_\varphi }} \right| = K{}_\varphi $. The learning task can be formulated as follows:
\begin{align}\label{learning task}
F\left( {\mathbf{w}} \right) = \frac{1}{K}\sum\nolimits_{k = 1}^K {{K_\varphi }{F_\varphi }\left( {{\mathbf{w}};{\mathcal{D}_\varphi }} \right)} ,
\end{align}
and
\begin{align}\label{loos fucnction}
{F_\varphi }\left( {{\mathbf{w}};{\mathcal{D}_\varphi }} \right) = \frac{1}{{{K_\varphi }}}\sum\nolimits_{\left( {{{\mathbf{x}}_{\varphi k}},{y_{\varphi k}}} \right) \in {\mathcal{D}_\varphi }} {f\left( {{\mathbf{w}};{{\mathbf{x}}_k},{y_k}} \right)} .
\end{align}
respectively.

Subsequently, FL is performed, in which edge devices work to minimize ${F_\varphi }$ to derive the local model, while the BS consolidates the global model based on the local updates. In particular, the model ${\mathbf{w}}$ is updated iteratively over $T$ training rounds. The $t$-th training round encompasses the following steps
\subsubsection{Global Broadcast}
Initially, the BS disseminates the most recent system model ${\mathbf{w}}$ to all users.
\subsubsection{Calculation of the Global Gradient}
Each edge device calculates its local gradient based on the local dataset. Specifically, the gradient can be expressed as
\begin{align}\label{locel gradent}
{{\mathbf{g}}_{\varphi ,t}} = \nabla {F_\varphi }\left( {{{\mathbf{w}}_t};{\mathcal{D}_\varphi }} \right) \in {\mathbb{R}^D},
\end{align}
where $\nabla {F_\varphi }\left( {{{\mathbf{w}}_t}; {\mathcal{D}_\varphi }} \right)$ is the gradient of $\nabla {F_\varphi }\left( \bullet \right)$ at ${\mathbf{w}} = {{\mathbf{w}}_t}$.
\subsubsection{Model Aggregation}
The edge device sends the set of gradients $\left\{ {{{\mathbf{g}}_{\varphi ,t}}} \right\}$ o the BS via the wireless channel. Upon receiving these signals, the BS updates the global gradient by computing a weighted sum of the gradients $\left\{ {{{\mathbf{g}}_{\varphi ,t}}:\varphi  \in \Phi } \right\}$. In this paper, we adopt the widely accepted FL framework, where the weight of the $\varphi $-th local gradient is proportional to its corresponding local dataset ${F_\varphi }$. In this case, we hope to estimate the global gradient vector in the received signal: ${\mathbf{r}} \triangleq \sum\nolimits_{\varphi  \in \Phi } {{K_\varphi }{G_{\varphi ,t}}} $ and using ${\mathbf{\hat r}}$ to denote the estimate of ${\mathbf{r}}$. The estimates ${\mathbf{\hat r}}$ are inevitably distorted due to channel fading and communication noise, so usually ${\mathbf{\hat r}} \ne {\mathbf{r}}$. After obtaining ${\mathbf{\hat r}}$, we can update ${{\mathbf{w}}_{t + 1}}$ by gradient descent algorithm, which can be written as
\begin{align}\label{gradient descent algorithm}
{{\mathbf{w}}_{t + 1}} = {{\mathbf{w}}_t} - \frac{\eta }{{\sum\nolimits_{\varphi  \in \Phi }{{K_\varphi }} }}{\mathbf{\hat r}},
\end{align}
where $\frac{\eta }{{\sum\nolimits_{\varphi  \in \Phi } {{K_\varphi }} }}$ is a scaling factor and $\eta $ is the learning rate.

To simplify the computation, it is assumed that the wireless channels are invariant during the learning process and all satisfy the quasi-static smooth fading model.
\subsection{FL Model Aggregation}
In order to reduce the amount of information uploaded during transmission, and in order to avoid leakage of the amount of user information, we only upload the gradient of each user when uploading information. Specifically, during the $t$-th training round, the edge devices utilize the identical time-frequency resources to transmit their local gradient values  $\left\{ {{{\mathbf{g}}_\varphi }:\varphi  \in \Phi } \right\}$. For simplicity of representation, ${{\mathbf{g}}_\varphi }$ is used to denote the gradient value of the $\varphi $-th user, and $\left\{ {{g_\varphi }\left[ d \right]:1 \leqslant d \leqslant D} \right\}$ is used to denote the $d$-th term of ${{\mathbf{g}}_\varphi }$. To convert the signal from a distributed device into a gradient, we begin by calculating the local gradient statistics, effectively transforming the gradient into a scalar value, i.e., the variance and the mean
${\bar g_\varphi } = \frac{1}{D}{\sum\nolimits_{d = 1}^D g _\varphi }\left[ d \right]$, $v_\varphi ^2 = \frac{1}{D}\sum\nolimits_{d = 1}^D {\left( {{g_\varphi }\left[ d \right] - {{\bar g}_\varphi }} \right)}$.
The next step is to upload that scalar to BS. To avoid the effects of different scales of eigenvalues, the data were preprocessed using zero-mean unit variance, i.e.,  map ${{\mathbf{g}}_\varphi }$ to ${x_\varphi }$ using ${\bar g_\varphi }$ and ${v_\varphi }$. At this point, the transmit sequence $\left\{ {{x_\varphi }:\varphi  \in \Phi } \right\}$ of the $\varphi $-th user can be represented as
\begin{align}\label{transmit sequence}
{x_\varphi } \triangleq \frac{{{g_\varphi }\left[ d \right] - {{\bar g}_\varphi }}}{{{v_\varphi }}}.
\end{align}
$\mathbb{E}\left[ {{x_\varphi }\left[ d \right]} \right] = 0$ and $\mathbb{E}\left[ {x_\varphi ^2\left[ d \right]} \right] = 1$ is satisfied by ${x_\varphi }$, when normalising we should also ensure that $\mathbb{E}\left[ {{{\left| {{p_\varphi }{x_\varphi }\left[ d \right]} \right|}^2}} \right] = {\left| {{p_\varphi }} \right|^2}$, where is ${{p_\varphi }}$ is the transmit power of ${U_\varphi }$.

Substituting the updated (\ref{transmit sequence}) into (\ref{BS receive signal}), let ${s_{n\Delta }} = \frac{{\sqrt {{p_n}} }}{{{v_n}}}\left( {{g_n}\left[ d \right] - {{\bar g}_n}} \right)$ and ${s_{m\Delta }} = \frac{{\sqrt {{p_m}} }}{{{v_m}}}\left( {{g_m}\left[ d \right] - {{\bar g}_m}} \right)$, at which point the user-set $n$ and user-set $m$ can be equated as user $n$ and user $m$. The expression for the received signal in round $t$ can be written as
\begin{align}\label{receivce single}
{y_t}& = {{\mathbf{f}}^H}\sum\nolimits_{n = 1}^N {({{\mathbf{h}}_{sn}} + {\mathbf{h}}_{sr}^H{{\mathbf{\Theta }}^n}{{\mathbf{h}}_{rn}})} {s_{n\Delta }}+{{\mathbf{f}}^H}{\mathbf{h}}_{sr}^H{{\mathbf{\Theta }}^n}{{\mathbf{n}}_s}\nonumber \\
 &+ {{\mathbf{f}}^H}{\mathbf{h}}_{sr}^H{{\mathbf{\Theta }}^m}\sum\nolimits_{m = 1}^M {{{\mathbf{h}}_{rm}}{s_{m\Delta }}}+ {{\mathbf{f}}^H}{\mathbf{h}}_{sr}^H{{\mathbf{\Theta }}^m}{{\mathbf{n}}_s}  + {\tilde n_0},
\end{align}
where ${\mathbf{f}  \in \mathbb{C}^{J \times 1}}$ epresents the normalized receiver beamforming vector, satisfying the condition $\left\| {\mathbf{f}} \right\|_2^2 = 1$.
The BS initially decodes the signal from user $n$ with favorable channel conditions. Subsequently, SIC is applied to subtract the signal from user $n$, followed by the decoding of the signal from user $m$.

The BS calculates the estimate of ${{\hat r}}$ from ${{ y}_t}$ using a linear estimator, which can be expressed as follows
\begin{align}\label{linear estimator}
{{\hat r_t}} = \frac{1}{{\sqrt u }}{{ y}_t} + \bar g,
\end{align}
where $\bar g \triangleq \sum\nolimits_{\varphi  \in \Phi } {{K_\varphi }} {\bar g_\varphi }$. $u > 0$ is a normalization scalar. Since ${\bar g_m}$ disappears in the normalisation process, the additional term $\bar g$ associated with it is introduced here. Each time a gradient upload is performed, the users' positions are assumed to remain fixed. \footnote{It is important to address dynamic user distributions in network operations for the purpose of practical scenarios, which will be discussed in our further works.}
\section{FL Performance Analysis and Problem Formulation}
In this section, we analyze the learning performance of ASTARS uplink networks utilizing FL, providing an upper bound on the learning loss along with a closed-form solution. This issue is framed as a task aimed at minimizing the learning loss.
\subsection{Problematic Assumption}
The estimate of the global gradient vector ${{\bf{\hat r}}_t} = \left[ {{{\hat r}_1}, \cdots ,{{\hat r}_t}} \right]$ is provided by \eqref{linear estimator}, where ${{\bf{e}}_t}$ the error is defined as the difference between the estimate and the true value. Hence the gradient descent formula is updated as
\begin{align}\label{global model update}
{{\bf{w}}_{t + 1}} = {{\bf{w}}_t} - \frac{\eta }{{\sum\nolimits_{\varphi  \in \Phi } {{K_\varphi }} }}{{\bf{\hat r}}_t} = {{\bf{w}}_t} - \eta \left( {\nabla F\left( {{{\bf{w}}_t}} \right) - {{\bf{e}}_t}} \right),
\end{align}
where $\nabla F\left( {\mathbf{w}} \right) \buildrel \Delta \over = \frac{1}{K}\sum\nolimits_{k = 1}^K {f\left( {{\mathbf{w}};{{\mathbf{x}}_k},{y_k}} \right)}$ represents the gradient of $F\left( {\mathbf{w}} \right)$ evaluated at ${\mathbf{w}} = {{\mathbf{w}}_t}$, and ${{\mathbf{e}}_t}$ denotes the gradient error vector arising from channel noise and model aggregation. The expression for ${{\bf{e}}_t}$ is given by
\begin{align}\label{gradient error}
{{\bf{e}}_t} = \underbrace {\nabla F({{\bf{w}}_t}) - \frac{{{{\bf{r}}_t}}}{{\sum\limits_{\varphi  \in \Phi } {{K_\Phi }} }}}_{{{\bf{e}}_{1,t}}} + \underbrace {\frac{{{{\bf{r}}_t}}}{{\sum\limits_{\varphi  \in \Phi } {{K_\varphi }} }} - \frac{{{{\widehat {\bf{r}}}_t}}}{{\sum\limits_{\varphi  \in \Phi } {{K_\varphi }} }}}_{{{\bf{e}}_{2,t}}},
\end{align}
where ${{\bf{e}}_{1,t}}$ is  the difference between the gradient of $\nabla F({\bf{w}})$ at ${{\bf{w}}_t}$ and the true gradient. This error arises from the fact that, due to channel fading, only part of the device's gradient signal is received at the BS, introducing an error with the true gradient value. At this point where $\left| \Phi  \right| < \Phi $, we employ ASTARS to amplify the transmit signal power of the distribution device, thereby minimizing this error to the maximum extent possible, i.e., ensuring that $\Phi  = \left\{ {1,2, \cdots ,\Phi } \right\}$. The term ${{\bf{e}}_{2,t}}$ is caused by noise, which cannot be avoided but can be minimized by adjusting variables such as $\left\{ {{\bf{f}},{\bf{\Theta }}^r,{\bf{\Theta }}^t ,\Phi ,{p_\varphi }} \right\}$.

For ease of handling, refer to the setup of $F\left( {\bf{w}} \right)$ in \cite{math2012hybrid} and \cite{konar2005neuro}:
\subsubsection{Strongly convex function}\label{a}
The global loss function $F({\bf{w}})$ exhibits strong convexity characterized by a positive parameter $\rho $. Therefore, for any ${\bf{w}}$ and ${\bf{v}}$,  the following inequality holds
\begin{align}\label{F strongly convex}
F\left( {\mathbf{w}} \right) \geqslant& F\left( {\mathbf{v}} \right) + {\left( {{\mathbf{w}} - {\mathbf{v}}} \right)^T}\nabla F\left( {\mathbf{v}} \right) \nonumber \\
&+ \frac{\rho }{2}\left\| {{\mathbf{w}} - {\mathbf{v}}} \right\|_2^2,\forall \mathbf{w},\mathbf{v} \in {\mathbb{R}^{D \times 1}}.
\end{align}
\subsubsection{Lipschitz continuum condition}\label{b}
The gradient $\nabla F\left( {\bf{w}} \right)$ satisfies the Lipschitz continuity condition with parameter $L$, i.e., for any ${\bf{w}}$ and ${\bf{v}}$, there exist
\begin{align}\label{lipschitz continuous}
{\left\| {\nabla F\left( {\mathbf{w}} \right) - \nabla F\left( {\mathbf{v}} \right)} \right\|_2} \leqslant L{\left\| {{\mathbf{w}} - {\mathbf{v}}} \right\|_2},\forall {\mathbf{w}},{\mathbf{v}} \in {\mathbb{R}^{D \times 1}}.
\end{align}
\subsubsection{Twice-continuously differentiable}\label{c}
$F\left( {\bf{w}} \right)$ is twice consecutively differentiable, which follows
\begin{align}\label{twice-continuously differentiable}
\rho {\mathbf{I}} \prec {\nabla ^2}F\left( {\mathbf{w}} \right) \prec L{\mathbf{I}}.
\end{align}
\subsubsection{Upper bound}\label{d}
Any $\nabla F\left(  \cdot \right)$ (gradient of the training sample) has an upper bound at $\left\{ {{w_t}:1 \le t \le T} \right\}$, i.e.,
\begin{align}\label{upper bonuded}
\left\| {f\left( {{\mathbf{w}};{{\mathbf{x}}_k},{y_k}} \right)} \right\| \leqslant {\alpha _1} + {\alpha _2}\left\| {\nabla F\left( {{\bf{w}}_t} \right)} \right\|_2^2,1 \leqslant k \leqslant K,
\end{align}
where ${\alpha _1}$ and ${\alpha _2}$ are constants that satisfy ${\alpha _1} \ge 0$ and ${\alpha _2} \ge 0$.

Equation \eqref{F strongly convex} guarantees the existence of an optimal solution to the loss function. The set of equations from \eqref{F strongly convex} to \eqref{twice-continuously differentiable} are commonly used in the context of solving optimization problems. Meanwhile, equation \eqref{upper bonuded} provides an upper bound on the gradient for each sample, which is yet to be determined.
\begin{theorem} \label{theorem:2}
Given that $F\left( {{{\bf{w}}_t}} \right)$ satisfies \eqref{F strongly convex}-\eqref{upper bonuded}, the upper limit of $F\left( {{{\bf{w}}_{t+1}}} \right)$ can be derived as
\begin{align}\label{wt+1 upper limit}
\mathbb{E}\left[ {F({{\mathbf{w}}_{t + 1}})} \right] \leqslant \mathbb{E}\left[ {F({{\mathbf{w}}_t})} \right] - \frac{1}{{2L}}\left\| {\nabla F({{\mathbf{w}}_t})} \right\|_2^2 + \frac{1}{{2L}}\mathbb{E}\left[ {\left\| {{{\mathbf{e}}_t}} \right\|_2^2} \right]. 
\end{align}
\begin{proof}
See Appendix~A.
\end{proof}
\end{theorem}

\subsection{Learning Performance}
Having established above that $F\left(  \cdot  \right)$ satisfies conditions \eqref{F strongly convex}-\eqref{upper bonuded} and Theorem 1, the analysis of learning performance is presented next.
Since ${{\bf{e}}_t}$ can be decomposed into ${{\bf{e}}_{1,t}}$ and ${{\bf{e}}_{2,t}}$, and the value of the device loss error ${{\bf{e}}_{1,t}}$ is extremely small or considered to be zero after augmenting the signals from the distribution device using ASTARS, \eqref{wt+1 upper limit} can be rewritten as
\begin{align}\label{more simplify F(w+1)}
&\mathbb{E}\left[ {F({{\mathbf{w}}_{t + 1}})} \right] \nonumber \\
&\leqslant \mathbb{E}\left[ {F({{\mathbf{w}}_t})} \right] - \frac{1}{{2L}}\left\| {\nabla F({{\mathbf{w}}_t})} \right\|_2^2 + \frac{1}{{2L}}\mathbb{E}\left[ {\left\| {{{\mathbf{e}}_{2,t}}} \right\|_2^2} \right]. 
\end{align}

To analyze $\mathbb{E}\left[ {F({{\mathbf{w}}_{t + 1}})} \right]$, it is also necessary to obtain an expression for $\frac{1}{{2L}}\mathbb{E}\left[ {\left\| {{{\mathbf{e}}_{2,t}}} \right\|_2^2} \right]$. If $\Phi $ is given, the value of $\mathbb{E}\left[ {\left\| {{{\mathbf{e}}_{2,t}}} \right\|_2^2} \right]$ is provided by the following lemma \ref{theorem:3}.
\begin{lemma} \label{theorem:3}
When the total number of devices $\Phi $ is given and the transmit power satisfies the condition of \eqref{average transmit power}, $\mathbb{E}\left[ {\left\| {{{\mathbf{e}}_{2,t}}} \right\|_2^2} \right]$ can be further calculated as
\begin{align}\label{e2,t}
\mathbb{E}\left[ {\left\| {{{\mathbf{e}}_{2,t}}} \right\|_2^2} \right] = \frac{{\Phi \left( {\tau \sigma _s^2 + \sigma _0^2} \right)}}{{{P_\varphi }{{\left( {\sum\nolimits_{\varphi  \in \Phi } {{K_\varphi }} } \right)}^2}}}\mathop {\max }\limits_{\varphi  \in \Phi } \frac{{K_\varphi ^2\nu _\varphi ^2}}{{{{\left| {{{\mathbf{f}}^H}{{\mathbf{h}}_n}} \right|}^2} + {{\left| {{{\mathbf{f}}^H}{{\mathbf{h}}_m}} \right|}^2}}}.
\end{align}
\end{lemma}
\begin{proof}
See Appendix~B.
\end{proof}
\begin{remark}
We observe that ${P_\varphi }$ is not a necessary variable for $\mathbb{E}\left[ {\left\| {{{\mathbf{e}}_{2,t}}} \right\|} \right]$ since ASTARS amplifies the signal. Consequently, the communication error is exclusively associated with the parameters $\left\{ {Q,{\mathbf{f}},{{\bf{\Theta }}^n},{{\bf{\Theta }}^m}} \right\}$. Therefore, the main task is to reduce the signal aggregation error caused by noise and equipment loss, so as to further improve the learning accuracy of FL. By controlling the ${\mathbf{f}}$, ${{\bf{\Theta }}^n}$ and ${{\bf{\Theta }}^m}$, the air aggregation of the signal can be accurately controlled.
\end{remark}
\begin{theorem} \label{theorem2}
For any $\left\{ {Q,{\mathbf{f}},{\mathbf{\Theta }^n},{\mathbf{\Theta }^m}} \right\}$ and $t = 0,1, \cdots ,T - 1$, the upper bound of $\mathbb{E}\left[ {F({{\mathbf{w}}_{t + 1}}) - F({{\mathbf{w}}^ \star })} \right]$ can be given by
\begin{align}\label{fw-f0}
\mathbb{E}&\left[ {F({{\mathbf{w}}_{t + 1}}) - F({{\mathbf{w}}^ \star })} \right] \nonumber \\
&\leqslant \frac{{{\alpha _1}}}{{2L}}d\left\{ {Q,{\mathbf{f}},{\mathbf{\Theta }^n},{\mathbf{\Theta }^m}} \right\}\frac{{1 - {{\left( {\Upsilon \left\{ {Q,{\mathbf{f}}, {\mathbf{\Theta }^n}, {\mathbf{\Theta }^m}} \right\}} \right)}^t}}}{{1 - \Upsilon \left\{ {Q,{\mathbf{f}},{\mathbf{\Theta }^n},{\mathbf{\Theta }^m}} \right\}}} \nonumber\\
 &+ {\left( {\Upsilon \left\{ {Q,{\mathbf{f}},{\mathbf{\Theta }^n},{\mathbf{\Theta }^m}} \right\}} \right)^t}\left( {F({{\mathbf{w}}_0}) - F({{\mathbf{w}}^ \star })} \right),
\end{align}
where $d\left\{ {Q,{\bf{f}},{\mathbf{\Theta }^n},{\mathbf{\Theta }^m}}\right\} \triangleq $$\frac{{\tau \sigma _s^2 + \sigma _0^2}}{{{P_\varphi }{{\left( {\sum\nolimits_{\varphi  \in \Phi } {{K_\varphi }} } \right)}^2}}}\mathop{\max }\limits_{\varphi  \in \Phi } \frac{{K_\varphi ^2}}{{{{\left| {{{\mathbf{f}}^H}{{\mathbf{h}}_n}} \right|}^2} + {{\left| {{{\mathbf{f}}^H}{{\mathbf{h}}_m}} \right|}^2}}}$ and $\Upsilon \left\{ {Q,{\mathbf{f}},{\mathbf{\Theta }^n},{\mathbf{\Theta }^m}} \right\} \triangleq \left\{ {1 - \frac{\rho }{L} + \frac{\rho }{L}{\alpha _2}d\left\{ {Q,{\mathbf{f}},{\mathbf{\Theta }^n},{\mathbf{\Theta }^m}} \right\}} \right\}$.
\end{theorem}
\begin{proof}
See Appendix~C.
\end{proof}
\begin{corollary}\label{T unlimite}
Suppose $d\left\{ {Q,{\mathbf{f}},{{\bf{\Theta }}^n},{{\bf{\Theta }}^m}} \right\} \leqslant \frac{1}{{{\alpha _2}}}$ holds whenever $\Upsilon \left\{ {Q,{\mathbf{f}},{{\bf{\Theta }}^n},{{\bf{\Theta }}^m}} \right\} < 1$ is satisfied. As $T \to \infty $, the optimal value $\mathbb{E}\left[ {F({{\mathbf{w}}_T}) - F({{\mathbf{w}}^ \star })} \right]$ can be calculated as
\begin{align}\label{t unlimite}
\mathop {\lim }\limits_{T \to \infty } \mathbb{E}\left[ {F({{\mathbf{w}}_T}) - F({{\mathbf{w}}^ \star })} \right] = \frac{{{\alpha _1}d\left\{ {Q,{\mathbf{f}},{{\bf{\Theta }}^n},{{\bf{\Theta }}^m}} \right\}}}{{2\rho \left( {1 + {\alpha _2}d\left\{ {Q,{\mathbf{f}},{{\bf{\Theta }}^n},{{\bf{\Theta }}^m}} \right\}} \right)}}.
\end{align}
\end{corollary}
\begin{remark}
The above formula indicates that when $d\left\{ {Q,{\mathbf{f}},{{\bf{\Theta }}^n},{{\bf{\Theta }}^m}} \right\}$ is small enough, $\mathbb{E}\left[ {F({{\mathbf{w}}_T}) - F({{\mathbf{w}}^ \star })} \right]$ can achieve convergence, and the rate of convergence is linear with $\Upsilon \left\{ {Q,{\mathbf{f}},{{\bf{\Theta }}^n},{{\bf{\Theta }}^m}} \right\}$. As $\Upsilon \left\{ {Q,{\mathbf{f}},{{\bf{\Theta }}^n},{{\bf{\Theta }}^m}} \right\}$ becomes smaller, $\mathbb{E}\left[ {F({{\mathbf{w}}_T}) - F({{\mathbf{w}}^ \star })} \right]$ converges faster. Since $\Upsilon \left\{ {Q,{\mathbf{f}},{{\bf{\Theta }}^n},{{\bf{\Theta }}^m}} \right\}$ is proportional to $d\left\{ {Q,{\mathbf{f}},{{\bf{\Theta }}^n},{{\bf{\Theta }}^m}} \right\}$, $d\left\{ {Q,{\mathbf{f}},{{\bf{\Theta }}^n},{{\bf{\Theta }}^m}} \right\}$ affects the learning performance. It is possible to use $d\left\{ {Q,{\mathbf{f}},{{\bf{\Theta }}^n},{{\bf{\Theta }}^m}} \right\}$  as a measure of the performance of FL and minimize $d\left\{ {Q,{\mathbf{f}},{{\bf{\Theta }}^n},{{\bf{\Theta }}^m}} \right\}$ by optimizing $\left\{ {Q,{\mathbf{f}},{{\bf{\Theta }}^n},{{\bf{\Theta }}^m}} \right\}$.
\end{remark}
\subsection{Problem Formulation}
It is worth pointing out that observation $d\left\{ {Q,{\mathbf{f}},{{\bf{\Theta }}^n},{{\bf{\Theta }}^m}} \right\}$ reveals that the values of $d\left\{ {Q,{\mathbf{f}},{{\bf{\Theta }}^n},{{\bf{\Theta }}^m}} \right\}$ are independent of $\left\{ {\rho ,L,{\alpha _1},{\alpha _2}} \right\}$. Thus, \eqref{t unlimite} holds for all general FL loss functions satisfying \eqref{F strongly convex}-\eqref{upper bonuded}.
It is not necessary to consider the effect of the values of the $\left\{ {\rho ,L,{\alpha _1},{\alpha _2}} \right\}$ variables when designing the optimization problem. The existence of an upper limit to this error, i.e., convergence feasibility, was verified by analyzing the error of $F\left( {{{\bf{w}}_t}} \right)$ with respect to the ideal gradient value $F\left( {{{\bf{w}}^*}} \right)$. The error was then analyzed, and the convergence feasibility verified. Subsequently, the upper limit of this error was analyzed, and the possibility of the error converging to 0 was verified after iteration. The next step is to consider how to optimize the error function so that it converges to the optimal gradient value $F\left( {{{\bf{w}}^*}} \right)$ as quickly and accurately as possible.
Since the number of reflectors in ASTARS is more challenging to adjust in reality, $\left\{ {{\mathbf{f}},{{\bf{\Theta }}^n},{{\bf{\Theta }}^m}} \right\}$ can be changed more flexibly. Therefore, $\left\{ {{\mathbf{f}},{{\bf{\Theta }}^n},{{\bf{\Theta }}^m}} \right\}$ are taken as the main optimization objectives. Upon analysis, we can reframe the optimization problem as follows
\begin{subequations}
\begin{align}\label{optimizationproblem1}
\mathop {\min }\limits_{{\bf{f}},{\mathbf{\Theta}^r},{\mathbf{\Theta}^t} }& \mathop {\max }\frac{{K_\varphi ^2}}{{{{\left| {{{\bf{f}}^H}{{\bf{h}}_n}} \right|}^2} + {{\left| {{{\bf{f}}^H}{{\bf{h}}_m}} \right|}^2}}},\\
{\rm{s}}{\rm{.t}}{\rm{.}}&{\mathbf{\Theta}}^n \in \left\{ {0,2\pi } \right\},{\mathbf{\Theta}}^m \in \left\{ {0,2\pi } \right\},\forall q \in Q,\\\
&{\beta _t} \in \left( {0,1} \right),{\beta _r} \in \left( {0,1} \right),{\beta _t} + {\beta _r} = 1,\\
&\left\| {\bf{f}} \right\|_2^2 = 1,\\
&\frac{1}{{N + M}}\sum\nolimits_{\varphi  = 1}^\Phi  {{{\left| {{p_\varphi }} \right|}^2}}  \leqslant {\bar P_\varphi }.
\end{align}
\end{subequations}
\subsection{Further Discussions}
If ${K_\varphi }$ is given, the \eqref{optimizationproblem1} is approximately concave, and we should perform some mathematical treatment to transform it into a convex or approximately convex function.
Observing $d\left\{ {Q,{\bf{f}},{{\bf{\Theta }}^n},{{\bf{\Theta }}^m}} \right\}$, it can be seen that the variables affecting the convergence of $F\left( {\bf{w}} \right)$ are: $\left\{ {\tau ,\sigma _s^2,{P_\varphi },{K_\varphi },{{\bf{f}}},{{\bf{h}}_n }, {{\bf{h}}_m }} \right\}$, while for a given amount of equipment and user transmit power, the remaining variables are constants, affected only by $\left\{ {{{\bf{f}}}, {{\bf{\Theta }}^n}, {{\bf{\Theta }}^m}} \right\}$.

We emphasize that the phase shift of ASTARS plays a critical role in optimizing learning performance, making accurate phase control essential for improving overall performance. Existing phase shift modes are typically categorized into two types: discrete phase shift and continuous phase shift. However, discrete phase shift inevitably introduces phase shift errors. In this paper, our primary focus is on enhancing FL performance through ASTARS. Therefore, we assume that ASTARS supports continuous phase-shift tuning, while considering discrete phase-shift as a direction for future research.
\section{Proposed Solutions}\label{Section_V}
In this section, we propose an iterative algorithm based on SCA to jointly optimize the received beamforming vector $\mathbf{f}$ and the ASTARS phase shift matrices ${{\mathbf{\Theta}}^n}$ and ${{\mathbf{\Theta}}^m}$. We also summarize the overall procedure and analyze its computational complexity.
\subsection{Optimizing Receiver Beamforming and RIS Phase Shifts}
It is assumed that the phases of the reflection and transmission elements of ASTARS are independent, exhibiting no coupling between them. Consequently, based on \eqref{optimizationproblem1}, the optimization problem can be articulated as follows:
\begin{align}\label{simply optimization}
\mathop {\min }\limits_{\left( {{\bf{f}},{{\bf{\Theta }}^n},{{\bf{\Theta }}^m}} \right) \in \Gamma } \mathop {\max } {\phi _{n,m}}\left( {{\bf{f}},{{\bf{\Theta }}^n},{{\bf{\Theta }}^m}} \right)& \notag \\
=& - \frac{{{{\left| {{{\bf{f}}^H}{{\bf{h}}_n}} \right|}^2} + {{\left| {{{\bf{f}}^H}{{\bf{h}}_m}} \right|}^2}}}{{K_\varphi ^2}},
\end{align}
where $\Gamma $ is described above. It is observed that the pronounced correlation between $\bf{f}$, ${{{\bf{\Theta }}^n}}$ and ${{{\mathbf{\Theta }}^m}}$ concerning $\beta $, coupled with the failure to mitigate this correlation through mathematical manipulation, results in the non-convex nature of \eqref{simply optimization}. Consequently, \eqref{simply optimization} constitutes a mixed integer non-convex optimization problem.

The SCA algorithm is employed to iteratively update $\bf{f}$, ${{{\bf{\Theta }}^n}}$ and ${{{\mathbf{\Theta }}^m}}$ by leveraging a sequence of approximate convex functions. First, we construct a proxy function $\tilde \phi _{n,m}^i\left( {{{\mathbf{f}}_i},{\mathbf{\Theta }}_i^n,{\mathbf{\Theta }}_i^m} \right)$, $i = 1,2, \cdots $, where $\left( {{{\mathbf{f}}_i},{\mathbf{\Theta }}_i^n,{\mathbf{\Theta }}_i^m} \right) \in \Gamma $.
At this juncture, $\tilde \phi _{n,m}^i\left( {{{\mathbf{f}}_i},{\mathbf{\Theta }}_i^n,{\mathbf{\Theta }}_i^m} \right)$ can be viewed as an approximate convex function of ....., and it can be expressed as
\begin{align}\label{SCA arg}
&\tilde \phi _{n,m}^i\left( {{{\bf{f}}_i},{\mathbf{\Theta }}_i^n,{\mathbf{\Theta }}_i^m} \right) \nonumber \\
&= {\phi _{n,m}}\left( {{{\bf{f}}_i},{\mathbf{\Theta }}_i^n,{\mathbf{\Theta }}_i^m} \right) + {\mathop{\rm Re}\nolimits} \left\{ {{{\left( {{\bf{f}} - {{\bf{f}}_i}} \right)}^H}{\nabla _{{{\bf{f}}_*}}}{\phi _{n,m}}\left( {{{\bf{f}}_i},{\mathbf{\Theta }}_i^n,{\mathbf{\Theta }}_i^m} \right)} \right\} \nonumber \\
&+ {\mathop{\rm Re}\nolimits} \left\{ {{{\left( {{{\bf{\Theta }}^n} - {\bf{\Theta }}_i^n} \right)}^H}{\nabla _{{\bf{\Theta }}_*^n}}{\phi _{n,m}}\left( {{{\bf{f}}_i},{\bf{\Theta }}_i^n,{\bf{\Theta }}_i^m} \right)} \right\} \nonumber \\
&+ {\mathop{\rm Re}\nolimits} \left\{ {{{\left( {{{\bf{\Theta }}^m} - {\bf{\Theta }}_i^m} \right)}^H}{\nabla _{{\bf{\Theta }}_*^m}}{\phi _{n,m}}\left( {{{\bf{f}}_i},{\bf{\Theta }}_i^n,{\bf{\Theta }}_i^m} \right)} \right\} \nonumber \\
&+ \frac{\varpi }{{{K^2}}}\left( {{{\left\| {{\bf{f}} - {{\bf{f}}_i}} \right\|}^2} + {{\left\| {{{\bf{\Theta }}^n} - {\bf{\Theta }}_i^n} \right\|}^2} + {{\left\| {{{\bf{\Theta }}^n} - {\bf{\Theta }}_i^m} \right\|}^2}} \right) \nonumber \\
&= \frac{\varpi }{{{K^2}}}\left( {{{\left\| {{\bf{f}} - {{\bf{f}}_i}} \right\|}^2} + {{\left\| {{{\bf{\Theta }}^n} - {\bf{\Theta }}_i^m} \right\|}^2} + {{\left\| {{{\bf{\Theta }}^n} - {\bf{\Theta }}_i^m} \right\|}^2}} \right) \nonumber \\
&+ \frac{{{{\left| {{{\bf{f}}^H}{{\bf{h}}_n}} \right|}^2} + {{\left| {{{\bf{f}}^H}{{\bf{h}}_m}} \right|}^2}}}{{{K^2}}} \\
&- \frac{2}{{{K^2}}}{\mathop{\rm Re}\nolimits} \left( {{{\bf{f}}^H}{{\bf{h}}_n}{{\bf{h}}_n}^H{{\bf{f}}_i} + {{\bf{f}}^H}{{\bf{h}}_m}{{\bf{h}}_m}^H{{\bf{f}}_i}} \right) \nonumber\\
&- \frac{2}{{{K^2}}}{\mathop{\rm Re}\nolimits} \left( {{{\bf{\Theta }}^n}{{\bf{h}}_{sr}}{{\bf{f}}_i}{{\left( {{{\bf{f}}_i}} \right)}^H}{{\bf{h}}_n}} \right) - \frac{2}{{{K^2}}}{\mathop{\rm Re}\nolimits} \left( {{{\bf{\Theta }}^m}{{\bf{h}}_{sr}}{{\bf{f}}_i}{{\left( {{{\bf{f}}_i}} \right)}^H}{{\bf{h}}_m}} \right) \nonumber\\
&+ \frac{2}{{{K^2}}}{\mathop{\rm Re}\nolimits} \left( {{{\left( {{{\bf{f}}_i}} \right)}^H}{{\bf{h}}_n}{{\bf{h}}_n}^H{{\bf{f}}_i} + {{\left( {{{\bf{f}}_i}} \right)}^H}{{\bf{h}}_m}{{\bf{h}}_m}^H{{\bf{f}}_i}} \right) \nonumber\\
&+ \frac{2}{{{K^2}}}{\mathop{\rm Re}\nolimits} \left( {{\bf{\Theta }}_i^n{{\bf{h}}_{sr}}{{\bf{f}}_i}{{\left( {{{\bf{f}}_i}} \right)}^H}{{\bf{h}}_n}} \right) + \frac{2}{{{K^2}}}{\mathop{\rm Re}\nolimits} \left( {{\bf{\Theta }}_i^m{{\bf{h}}_{sr}}{{\bf{f}}_i}{{\left( {{{\bf{f}}_i}} \right)}^H}{{\bf{h}}_m}} \right)\nonumber,
\end{align}
where ${\mathop{\rm Re}\nolimits} \left(  \bullet  \right)$ represents the real part of the operation, and ${\nabla _{{{\bf{f}}_*}}}{\phi _{n,m}}\left( {{{\bf{f}}_i},{\bf{\Theta }}_i^n,{\bf{\Theta }}_i^m} \right)$, ${\nabla _{{\bf{\Theta }}_*^n}}{\phi _{n,m}}\left( {{{\bf{f}}_i},{\bf{\Theta }}_i^n,{\bf{\Theta }}_i^m} \right)$ and ${\nabla _{{\bf{\Theta }}_*^m}}{\phi _{n,m}}\left( {{{\bf{f}}_i},{\bf{\Theta }}_i^n,{\bf{\Theta }}_i^m} \right)$ denotes the conjugate gradient of ${\phi _{n,m}}\left( {{{\bf{f}}_i},{\bf{\Theta }}_i^n,{\bf{\Theta }}_i^m} \right)$ at ${{\bf{f}}_i}$, ${\bf{\Theta }}_i^n$ and ${\bf{\Theta }}_i^m$. A suitable $\varpi $ is selected to ensure the strong convexity of $\tilde \phi _{n,m}^i\left( {{{\bf{f}}_i},{\bf{\Theta }}_i^n,{\bf{\Theta }}_i^m} \right)$.

${{\bf{f}}_{i + 1}}$, ${\bf{\Theta }}_{i + 1}^n$ and ${\bf{\Theta }}_{i + 1}^m$ are derived from $\left( {{{\bf{f}}_{i + 1}},{\bf{\Theta }}_{i + 1}^n,{\bf{\Theta }}_{i + 1}^m} \right) = \arg \mathop {\min }\limits_{\left( {{\bf{f,}}{{\bf{\Theta }}^n},{{\bf{\Theta }}^m}} \right) \in \Gamma } \mathop {\max }\limits_{n,m \in \Phi } {\phi _{n,m}}\left( {{\bf{f}},{{\bf{\Theta }}^n},{{\bf{\Theta }}^m}} \right)$. Since $\left\| {\bf{f}} \right\|_2^2 = 1$ for $\forall \left( {{\bf{f}},{{\bf{\Theta }}^n},{{\bf{\Theta }}^m}} \right) \in \Gamma $, \eqref{SCA arg} can be rewritten as
\begin{subequations}
\begin{align}\label{SCA2}
\left( {{{\bf{f}}_{i + 1}},{\mathbf{\Theta}} _{i + 1}^n,{\mathbf{\Theta}} _{i + 1}^m} \right) &= \arg \mathop {\min }\limits_{\left( {{\bf{f,}}{{\mathbf{\Theta}}^n},{{\mathbf{\Theta}}^m}} \right) \in \Gamma } \Lambda, \\
{\rm{s}}{\rm{.t}}{\rm{.}}K_\varphi ^2\Lambda \ge a_\varphi ^i - 2{\mathop{\rm Re}\nolimits}& \left\{ {{{\bf{f}}^H}{\bf{b}}_\varphi ^i + {\mathbf{\Theta}} _n^H{\bf{c}}_\varphi ^i + {\mathbf{\Theta}} _m^H{\bf{d}}_\varphi ^i} \right\}.
\end{align}
\end{subequations}
where $a_\varphi ^i = 2{\mathop{\rm Re}\nolimits} \left\{ {{{\left( {{{\bf{f}}_i}} \right)}^H}{{\bf{h}}_n}{{\bf{h}}_n}{{}^H}{{\bf{f}}_i} + {{\left( {{{\bf{f}}_i}} \right)}^H}{{\bf{h}}_m}{{\bf{h}}_m}{{}^H}{{\bf{f}}_i}} \right\}+ 2{\mathop{\rm Re}\nolimits} \left\{ {{\bf{\Theta}} _i^n{{\bf{h}}_{sr}}{{\bf{f}}_i}{{\left( {{{\bf{f}}_i}}\right)}^H}{{\bf{h}}_n} + {\bf{\Theta}} _i^m{{\bf{h}}_{sr}}{{\bf{f}}_i}{{\left( {{{\bf{f}}_i}} \right)}^H}{{\bf{h}}_m}} \right\}+{\left| {{{\bf{f}}^H}{{\bf{h}}_n}} \right|^2} + {\left| {{{\bf{f}}^H}{{\bf{h}}_m}} \right|^2} + 2\varpi \left( {L + 2} \right)$, ${\bf{b}}_\varphi ^i = \left( {\varpi  +{{\bf{h}}_n}{{\bf{h}}_n}{{}^H} + {{\bf{h}}_m}{{\bf{h}}_m}{{}^H}} \right){{\bf{f}}_i}$, ${\bf{c}}_\varphi ^i = \varpi {\bf{\Theta}} _i^n + {{\bf{h}}_{sr}}{{\bf{f}}_i}{\left( {{{\bf{f}}_i}} \right)^H}{{\bf{h}}_n}$ and ${\bf{d}}_\varphi ^i = \varpi {\bf{\Theta}} _i^m + {{\bf{h}}_{sr}}{{\bf{f}}_i}{\left( {{{\bf{f}}_i}} \right)^H}{{\bf{h}}_m}$.

The non-convexity of $\left( {{\bf{f}},{{\bf{\Theta }}^n},{{\bf{\Theta }}^m}} \right) \in \Gamma $ contributes to the persistence of non-convexity in problem \eqref{SCA2}. To address this challenge, we explore solving the Lagrangian dual problem to approximate the solution. Since there are $\Phi $ constraints, we define $\zeta  = \left| {{\zeta _1}, \cdots ,{\zeta _\varphi }, \cdots ,{\zeta _\Phi }} \right| \ge 0$ as the Lagrange multiplier vector. At this point, we introduce the Lagrangian pairwise function, defined as \eqref{lagelangri}.
\begin{figure*}[t]
{\small
\begin{align}\label{lagelangri}
&{{\cal L}_i}\left( {{\bf{f}},{{\bf{\Theta }}^n},{{\bf{\Theta }}^m},\Lambda } \right)= \Lambda  + \sum\limits_{\varphi  = 1}^\Phi  {{\zeta _\varphi }\left\{ {a_\varphi ^i - 2{\mathop{\rm Re}\nolimits} \left[ {{{\bf{f}}^H}{\bf{b}}_\varphi ^i + {{\left( {{{\bf{\Theta }}^n}} \right)}^H}{\bf{c}}_\varphi ^i + {{\left( {{{\bf{\Theta }}^m}} \right)}^H}{\bf{d}}_\varphi ^i} \right] - {K^2}\Lambda } \right\}} \\ \nonumber
& = \sum\limits_{\varphi  = 1}^\Phi  {{\zeta _\varphi }} a_\varphi ^i + \left( {1 - \sum\limits_{\varphi  = 1}^\Phi  {{\zeta _\varphi }{K^2}} } \right)\Lambda - 2{\mathop{\rm Re}\nolimits} \left\{ {{{\bf{f}}^H}\left( {\sum\limits_{\varphi  = 1}^\Phi  {{\zeta _\varphi }{\bf{b}}_\varphi ^i} } \right) + {{\left( {{{\bf{\Theta }}^n}} \right)}^H}\left( {\sum\limits_{\varphi  = 1}^\Phi  {{\zeta _\varphi }{\bf{c}}_\varphi ^i} } \right) + {{\left( {{{\bf{\Theta }}^m}} \right)}^H}\left( {\sum\limits_{\varphi  = 1}^\Phi  {{\zeta _\varphi }{\bf{d}}_\varphi ^i} } \right)} \right\} \nonumber.
\end{align}
\hrulefill
}
\end{figure*}
Transforming the dyadic problem $\mathop {\max }\limits_{\zeta  \geqslant 0} \mathop {\min }\limits_{\left( {{\mathbf{f,}}{{\mathbf{\Theta }}^n},{{\mathbf{\Theta }}^m}} \right) \in \Gamma ,\Lambda  \in \mathbb{R}} {\mathcal{L}_i}\left( {{\mathbf{f}},{{\bf{\Theta }}^n},{{\bf{\Theta }}^m},\Lambda ,\zeta } \right)$ into $\left( {{{\mathbf{f}}_{i + 1}},{\mathbf{\Theta }}_{i + 1}^n,{\mathbf{\Theta }}_{i + 1}^m} \right)$, ${{\mathbf{f}}_{i + 1}}(\zeta )$, ${\mathbf{\Theta }}_{i + 1}^n(\zeta )$ and ${\mathbf{\Theta }}_{i + 1}^m(\zeta )$ are accomplished by
\begin{subequations}
\begin{align}
{{\bf{f}}_{i + 1}}(\zeta ) &= \frac{{\sum\limits_{\varphi  \in \Phi } {{\zeta _\varphi }} {\bf{b}}_\varphi ^i}}{{{{\left\| {\sum\limits_{\varphi  \in \Phi } {{\zeta _\varphi }} {\bf{b}}_\varphi ^i} \right\|}_2}}},\label{f}\\
{\bf{\Theta }}_{i + 1}^n(\zeta )= &\sqrt {\lambda {\beta _n}}{\text{diag}}\left( {e^{j\arg \left\{ {\sum\limits_{\varphi  = 1}^\Phi  {{\zeta _\varphi }{\bf{c}}_\varphi ^i} } \right\}}}\right),\label{thetar}\\
{\bf{\Theta }}_{i + 1}^m(\zeta )= &\sqrt {\lambda {\beta _m}}{\text{diag}}\left( {e^{j\arg \left\{ {\sum\limits_{\varphi  = 1}^\Phi  {{\zeta _\varphi }{\bf{d}}_\varphi ^i} } \right\}}}\right) \label{thetat},
\end{align}
\end{subequations}
where $j = \sqrt { - 1} $. $\zeta $ is obtained by solving this convex problem
\begin{subequations}
\begin{align}\label{xi}
\mathop {\arg \min } \limits_\zeta  &2{\left\| {\sum\limits_{\varphi  = 1}^\Phi  {{\zeta _\varphi }{\bf{b}}_\varphi ^i} } \right\|_2} + 2{\left\| {\sum\limits_{\varphi  = 1}^\Phi  {{\zeta _\varphi }{\bf{c}}_\varphi ^i} } \right\|_1}+ 2{\left\| {\sum\limits_{\varphi  = 1}^\Phi  {{\zeta _\varphi }{\bf{d}}_\varphi ^i} } \right\|_1} \nonumber \\
&- \sum\limits_{\varphi  = 1}^\Phi  {{\zeta _\varphi }} a_\varphi ^i,\\
{\rm{s}}{\rm{.t}}{\rm{.}}\zeta  &\ge 0,\sum\limits_{\varphi  = 1}^\Phi  {{K^2}{\zeta _\varphi }}  = 1.
\end{align}
\end{subequations}
\begin{algorithm}
\caption{SCA Optimization Algorithm}\label{euclid}
\begin{algorithmic}[1]
\STATE \textbf{Input}: ${\bf{x}}$; $\varpi $; ${I_{\max }}$; $\xi $.
\STATE \textbf{Initialisation}: ${{\bf{f}}_1}$, ${\bf{\Theta }}_{1}^n$ and ${\bf{\Theta }}_{1}^m$.
\STATE Compute ${\rm{ob}}{{\rm{j}}_1} = \mathop {\min }\limits_{\varphi  \in \Phi }  - \frac{{ {{{\left| {{\bf{f}}_1^H{{\bf{h}}_n}} \right|}^2} + {{\left| {{\bf{f}}_1^H{{\bf{h}}_m}} \right|}^2}} }}{{{K^2}}}$
\STATE \textbf{for} $i = 1,2, \cdots ,{I_{\max }}$
\STATE Compute $\zeta $ by \eqref{xi};
\STATE Compute ${{\bf{f}}_{i + 1}},{\bf{\Theta }}_{i + 1}^n,{\bf{\Theta }}_{i + 1}^m$ by \eqref{f}, \eqref{thetar} and \eqref{thetat};
\STATE Update ${\rm{ob}}{{\rm{j}}_{i + 1}} = \mathop {\min }\limits_{\varphi  \in \Phi }  - \frac{{{{\left| {{\bf{f}}_{i + 1}^H{{\bf{h}}_n}} \right|}^2} + {{\left| {{\bf{f}}_{i + 1}^H{{\bf{h}}_m}} \right|}^2}}}{{{K^2}}}$
\STATE \textbf{if} $\frac{{\left| {{\rm{ob}}{{\rm{j}}_{i + 1}} - {\rm{ob}}{{\rm{j}}_i}} \right|}}{{\left| {{\rm{ob}}{{\rm{j}}_{i + 1}}} \right|}} \le \xi $, early end;
\STATE \textbf{end of}
\STATE \textbf{Output} ${{\bf{f}}_{i + 1}},{\bf{\Theta }}_{i + 1}^n,{\bf{\Theta }}_{i + 1}^m$.
\end{algorithmic}
\end{algorithm}

Since the original problem is non-convex, the results of problem \eqref{xi} are typically suboptimal. To avoid converging to a local optimum, the problem is solved using multiple initializations. Algorithm 1 outlines the optimization steps for  and ${{\mathbf{f}}_{i + 1}}(\zeta )$, ${\mathbf{\Theta }}_{i + 1}^n(\zeta )$ and ${\mathbf{\Theta }}_{i + 1}^m(\zeta )$. At this stage, given that \eqref{xi} represents a standard convex problem, we can employ a standard convex optimization solver to solve it. The algorithm terminates under two conditions: either the maximum number of iterations, denoted as ${I_{\max }}$ achieved, or the difference in objective values between successive iterations is less than a predefined threshold.

\begin{theorem} \label{theorem3}
Let ${\left\{ {{\phi _{n,m}}[k]} \right\}_k} \buildrel \Delta \over = {\left\{ {\left( {\phi _{n,m}^i[k]} \right)_{i = 1}^I} \right\}_k}$ denote the sequence generated by Algorithm 1, and define the averaged sequence as ${\left\{ {{{\bar \phi }_{n,m}}[k]} \right\}_k} \triangleq {\left\{ {\left( {1/I} \right)\sum\nolimits_{i = 1}^I {\phi _{n,m}^i[k]} } \right\}_k}$.  Assume that the step-size sequence ${\left\{ {\alpha [k]} \right\}_k}$ so that $\alpha [k] \in \left( {0,1} \right]$,  for all $k$, and
\begin{align}\label{eqtheorem3}
\sum\nolimits_{n = 0}^\infty  {\alpha [k] = \infty \qquad and \qquad \sum\nolimits_{n = 0}^\infty  {\alpha {{[k]}^2} \leqslant \infty } }.
\end{align}
Then the following statements hold:
\textit{a)} The averaged sequence  ${\left\{ {{{\bar \phi }_{n,m}}[k]} \right\}_k}$ is bounded, and all its limit points are stationary solutions of the problem in \eqref{simply optimization};
\textit{b)} The individual sequences ${\left\{ {\left( {\phi _{n,m}^i[k]} \right)} \right\}_k}$ are asymptotically consistent with the average, i.e., for all $i = 1, \ldots, I$, it holds that $\left\| {\phi _{n,m}^i[k] - {{\bar \phi }_{n,m}}[k]} \right\|\mathop  \to \limits_{n \to \infty } 0$.
\end{theorem}
\begin{proof}
See Appendix~D.
\end{proof}
\subsection{Algorithmic Complexity}
The computational complexity of each iteration (steps 5-8) in Algorithm~\ref{euclid} is dominated by solving a convex optimization problem using the interior point method, which has a per-iteration complexity of $O\left( {{{\left| \Phi  \right|}^3}} \right)$, where $\Phi $ represents the number of constraints in the optimization problem. Therefore, considering a maximum of $I_{max}$ iterations, the overall computational complexity of Algorithm~\ref{euclid} is upper bounded by  $O({I_{max} \Phi}^3)$. The space complexity is mainly determined by the size of optimization variables and parameters, approximately $O( Q^3)$. The computational complexity and space complexity of Algorithm~\ref{euclid} are lower than those of the MM algorithm and the DC algorithm. Additionally, the SCA algorithm typically exhibits linear or sublinear convergence behavior, usually converging in fewer iterations than $I_{max}$ in practical applications.

\begin{table}[htbp]
  \centering
  \caption{Simulation Parameters}
  \label{tab:sample}
  \begin{tabular}{|l|l|}
    \hline
    Monte Carlo simulation repeated & $10$ iterations  \\
    \hline
    Rician factor & $\kappa  =  - 5$ dB \\
    \hline
    Amplitude coefficients of ASTARS elements & ${\beta _r} = 0.7$, ${\beta _t} = 0.3$\\
    \hline
    Pass loss factors & $\alpha  = 2$, ${\eta _0} =  - 30$ dB \\
    \hline
    Amplified thermal noise & $\sigma_s^2=-70\mathrm{~dBm}$ \\
    \hline
    Noise power & $\sigma_0^2=\sigma_{re}^2=-90\mathrm{~dBm}$ \\
    \hline
    Learning rate & $\eta = 0.01$ \\
    \hline
    Number of users & $N=M=40$ \\
    \hline
    Regularization parameter & $\rho =1$ \\
    \hline
    Number of Reflection Units & $Q=30$  \\
    \hline
    Number of receiving antennas & $N=5$  \\
    \hline    
    Amplification Factors of ASTARS & $\lambda=5$  \\
    \hline
  \end{tabular}
\end{table}
\section{Simulation Results}
This section conducts simulation experiments to validate the accuracy of  performance analysis results in comparison with the performance of ASTARS uplink networks with FL. Additionally, the results are compared with various types of RIS. Unless stated otherwise, the simulation parameters are presented in Table \ref{tab:sample}. Consider a Cartesian 3D coordinate system with the BS located at $\left( {0,0,10} \right)$ and the ASTARS positioned at $\left( {50,0,10} \right)$. The set for local devices and $K$ are as follows: the $N$ devices in the reflective region are evenly distributed in region ${I_1} \buildrel \Delta \over = \left\{ {\left( {50 + x,50 + y,0} \right):0 \le x \le 20, - 10 \le y \le 10} \right\}$, and similarly the $M$  devices in the refractive region are evenly distributed in ${I_2} \buildrel \Delta \over = \left\{ {\left( {50 + x,50 + y,0} \right): - 20 \le x \le 0, - 10 \le y \le 10} \right\}$. Set 1 involves local users with identical data sizes, where $K$ is set to 750. In set 2, local users have varying data sizes, with half of the users' datasets randomly sampled in region $\left[ {100;200} \right]$ and the other half in region $\left[ {1000;2000} \right]$. Intuitively, the data from the different devices in set 1 are similar, indicating no significant discreteness issue. Conversely, in set 2, the data from various devices are heterogeneous, leading to a notable discreteness problem. 

To evaluate the performance of ASTARS uplink networks integrated with FL, we conduct experiments using the MNIST dataset. Specifically, we implement a convolutional neural network (CNN) for the handwritten digit classification task. The architecture consists of two convolutional layers with a kernel size of $5 \times 5$, each followed by a maximum pooling layer sized $2 \times 2$. This is followed by a batch normalization layer, a fully connected layer with 50 units, a ReLU activation layer, and a softmax output layer. The loss function utilized for training is the cross-entropy loss.

For the purpose of comparisons, we consider the following baselines in our experiments. All baselines are adjusted using the proposed SCA algorithm for $ {\bf{f}}$, ${{\bf{\Theta }}^n}$, and ${{\bf{\Theta }}^m}$:

\subsubsection{Noise-free} In light of the absence of communication noise throughout the communication process and the optimal reception and decoding of user messages, this scenario can be considered an ideal condition.
\subsubsection{PSTARS} In this case, the number of reflecting/transmitting elements is the same as in ASTARS. Nevertheless, due to the absence of signal amplification in PSTARS, signal attenuation may potentially result in data loss for a subset of users.
\subsubsection{RIS} Two RISs were implemented to accommodate both reflected and transmitted signals, thereby addressing the requirements of full-space users. However, this approach inevitably resulted in some signal loss.
\subsubsection{Without RIS} The user signal is transmitted directly to the BS, where the severe attenuation of the signal causes the BS to experience difficulties in receiving user data.
\subsection{FL with Batch Gradient Descent}
\begin{figure}[!h]
   \begin{center}
        \includegraphics[width=3.4in,  height=2.5in]{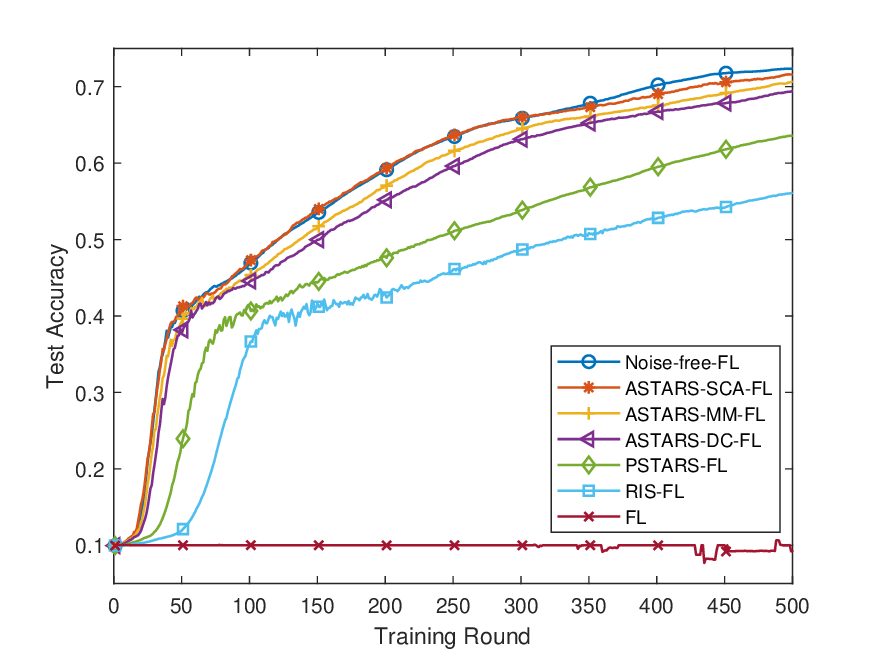}
        \caption{Test accuracy with set 1.}
        \label{Fig. 2}
    \end{center}
\end{figure}
\begin{figure}[!h]
    \begin{center}
        \includegraphics[width=3.4in,  height=2.5in]{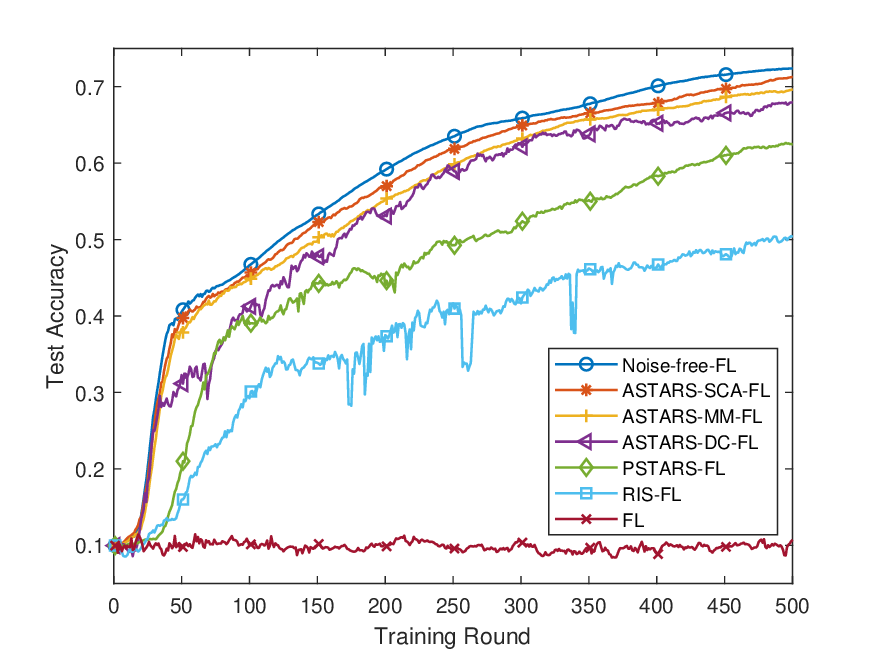}
        \caption{Text accuracy with set 2.}
        \label{Fig. 3}
    \end{center}
\end{figure}
In Fig. \ref{Fig. 2}, we plot the test accuracy of ASTARS uplink networks with FL for set 1, alongside the test accuracy for various comparison benchmarks. To prevent singular values, each curve is derived from ten Monte Carlo averages. As the data from different devices in set 1 are evenly distributed, there is no significant dispersion problem. Comparing the different types of RIS, it is evident that the ASTARS uplink networks with FL system exhibits the best performance. This illustrates the superiority of ASTARS in mitigating multiplicative fading through signal amplification. Additionally, by comparing the MM algorithm and the DC algorithm, the experiment found that both algorithms can also achieve higher test accuracy. This is due to ASTARS amplifying the uplink signal and mitigating the discreteness issue. Moreover, without the assistance of RIS, the performance of FL deteriorates. Similarly, Fig. \ref{Fig. 3} depicts the test accuracy of ASTARS uplink networks with FL for set 2. It is evident that the set 2 scenario encounters a significant discreteness issue, as reflected in the test accuracy. Similarly, Fig. \ref{Fig. 3} depicts the test accuracy of ASTARS uplink networks with FL for set 2. It is evident that the set 2 scenario encounters a significant discreteness issue, as reflected in the test accuracy.

\begin{figure}[!h]
    \begin{center}
        \includegraphics[width=3.4in,  height=2.5in]{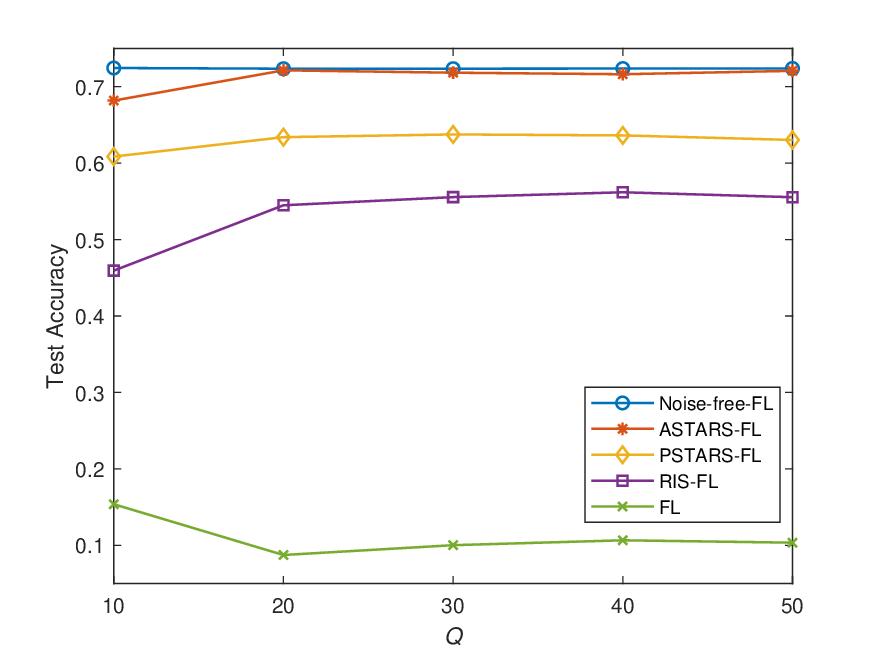}
      \caption{Test accuracy versus \emph{Q} with set 1.}
       \label{Fig. 4}
   \end{center}
\end{figure}
\begin{figure}[!h]
    \begin{center}
        \includegraphics[width=3.4in,  height=2.5in]{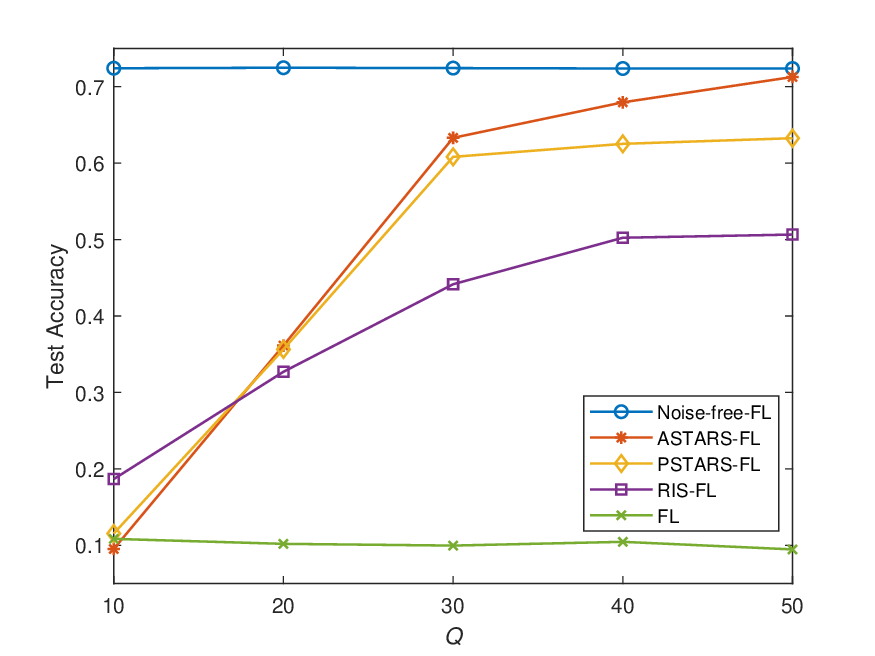}
        \caption{Test accuracy versus \emph{Q} with set 2.}
        \label{Fig. 5}
   \end{center}
\end{figure}

In Fig. \ref{Fig. 4} and \ref{Fig. 5}, we plot the test accuracy achieved with different numbers of reflection/transmission elements under the conditions of set 1 and set 2, respectively. Under the condition of set 1, various RISs configurations can attain the optimal test accuracy when $Q = 20$; under the condition of set 2, different RISs configurations can only approach the optimal test accuracy when $Q = 40$. ASTARS uplink networks with FL necessitates a smaller number of reflection/transmission elements to achieve optimal test accuracy compared to the findings reported in other literatures.
\subsection{FL with Mini-Batch Gradient Descent}
\begin{figure}[!h]
    \begin{center}
        \includegraphics[width=3.4in,  height=2.5in]{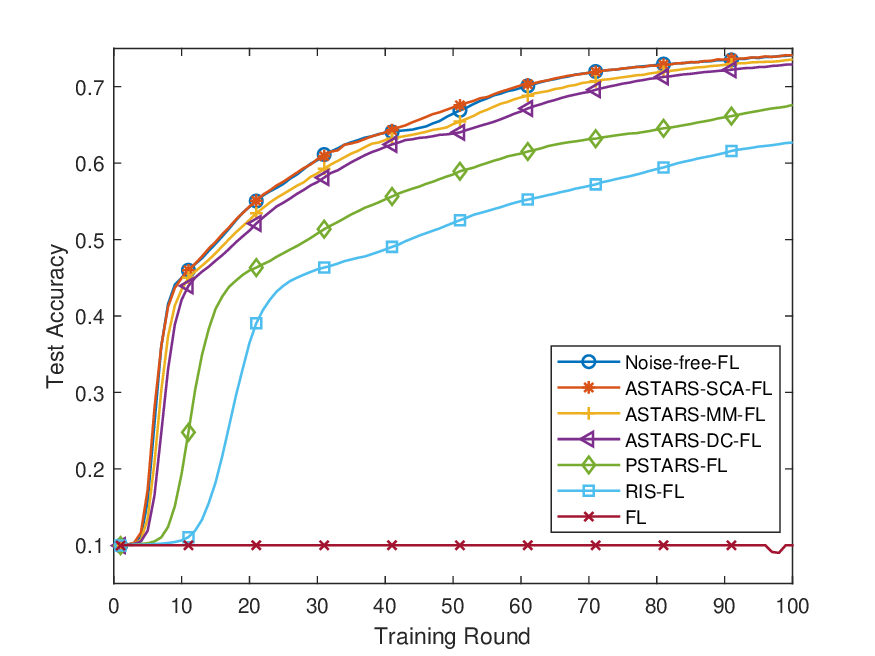}
        \caption{Mini-batch descent test accuracy with set 1.}
        \label{Fig. 6}
    \end{center}
\end{figure}
\begin{figure}[!h]
    \begin{center}
    \includegraphics[width=3.4in,  height=2.5in]{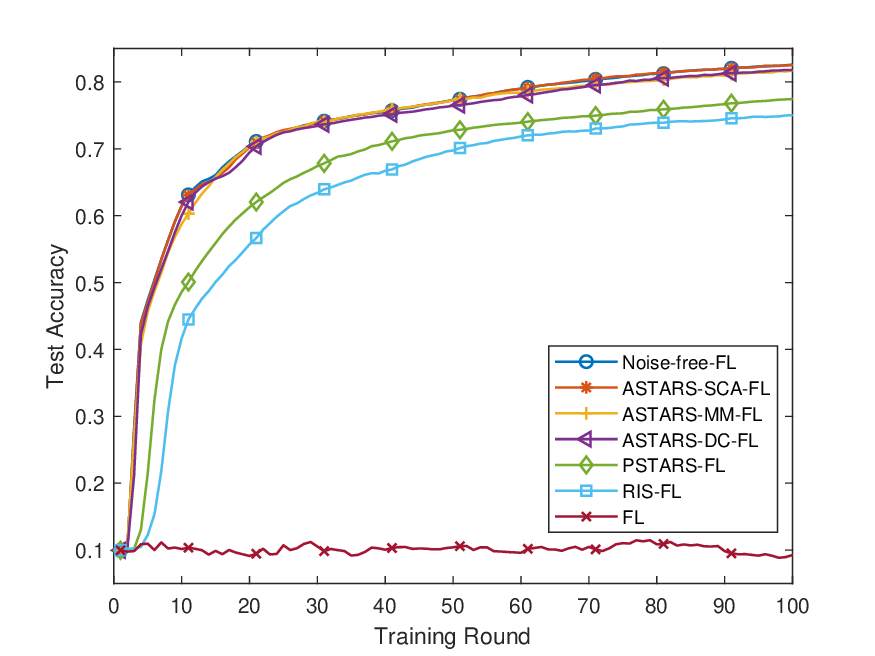}
        \caption{Mini-batch descent test accuracy with set 2.}
        \label{Fig. 7}
    \end{center}
\end{figure}
In this subsection, consider a FL framework for small-batch gradient descent. In particularly, the training dataset is first partitioned into small batches of equal size, the gradient $\eta _i^ * $ of each small batch is computed, then these gradients are averaged ${\eta ^ * } = \sum\nolimits_{i = 1}^I {{\raise0.7ex\hbox{${\eta _i^ * }$} \!\mathord{\left/
 {\vphantom {{\eta _i^ * } I}}\right.\kern-\nulldelimiterspace}
\!\lower0.7ex\hbox{$I$}}} $ at BS, and finally the above steps are repeated.

In Fig. \ref{Fig. 6}, we plot the test accuracy of ASTARS uplink networks with FL under set 1 and small-batch gradient descent. Additionally, it plots the test accuracy for different baselines. It is observed that the speed of FL convergence improves across all the different frameworks. This improvement is attributed to the fact that the FL system only needs to compute the gradient of each small batch during updating, rather than traversing the entire dataset. Calculating the gradient based on each small batch reduces the FL overhead, while the parameter updates are unaffected by singular values, resulting in higher test accuracy. Similarly, Fig. \ref{Fig. 7} illustrates the test accuracy of ASTARS uplink networks with FL under set 2 and small-batch gradient descent. The utilization of ASTARS uplink networks with FL mitigates the discreteness issue by reducing communication errors, thereby resulting in superior test accuracy compared to all baselines.
\begin{figure}[!h]
   \begin{center}
        \includegraphics[width=3.4in,  height=2.5in]{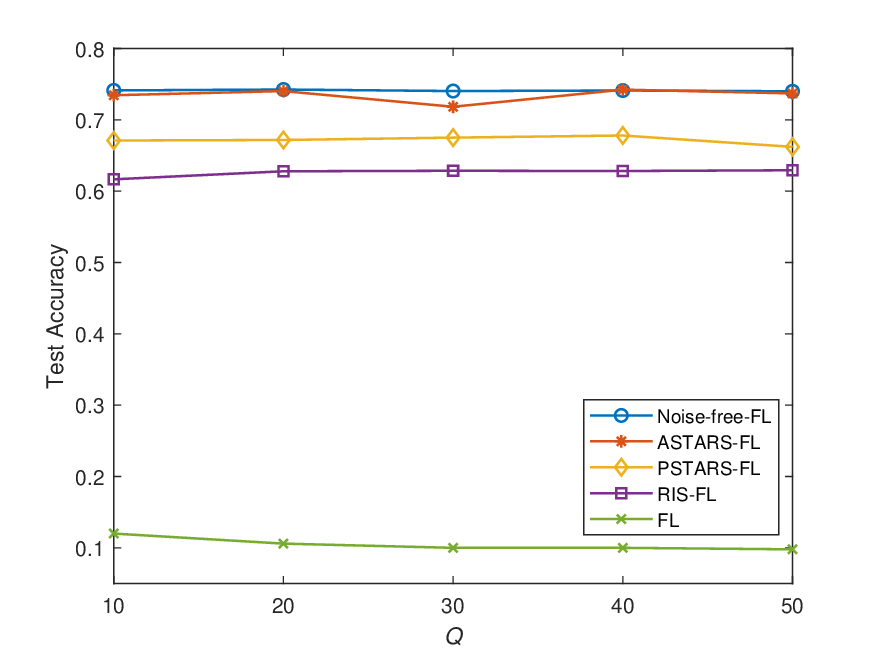}
       \caption{Mini-batch descent test accuracy versus \emph{Q} with set 1.}
       \label{Fig. 8}
    \end{center}
\end{figure}
\begin{figure}[!h]
    \begin{center}
        \includegraphics[width=3.4in,  height=2.5in]{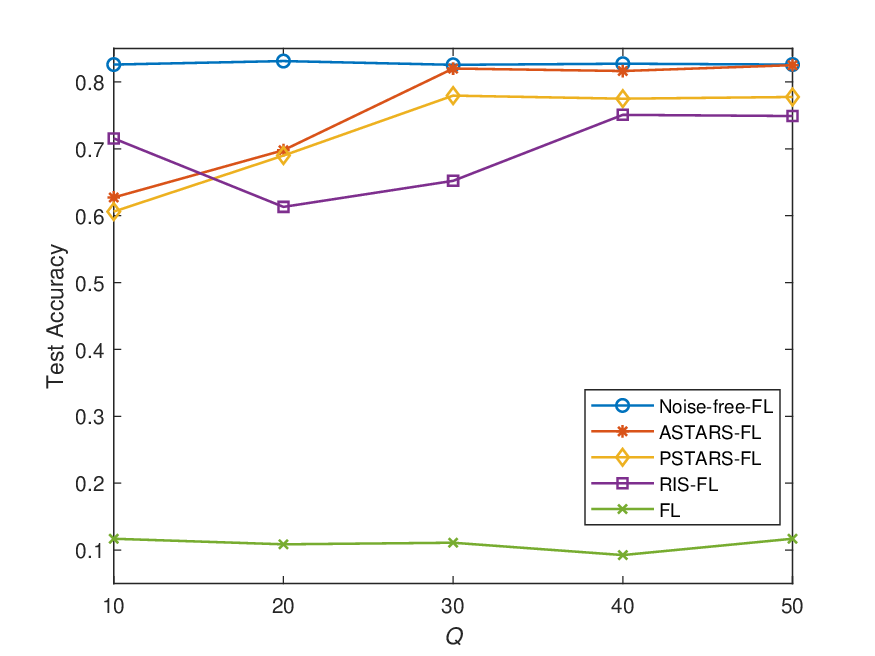}
        \caption{Mini-batch descent test accuracy \emph{Q} with set 2.}
        \label{Fig. 9}
    \end{center}
\end{figure}
Finally, Fig. \ref{Fig. 8} and \ref{Fig. 9} depict the test accuracy achieved with different numbers of reflection/transmission elements under the conditions of set 1 and set 2, respectively. It is observed that ASTARS uplink networks with FL can achieve the highest test accuracy. Additionally, ASTARS uplink networks requires fewer reflection units to achieve higher test accuracy under the condition of set 2.
\subsection{FL with Different Amplification Factors $\lambda$}
\begin{figure}[!h]
    \begin{center}
        \includegraphics[width=3.4in,  height=2.5in]{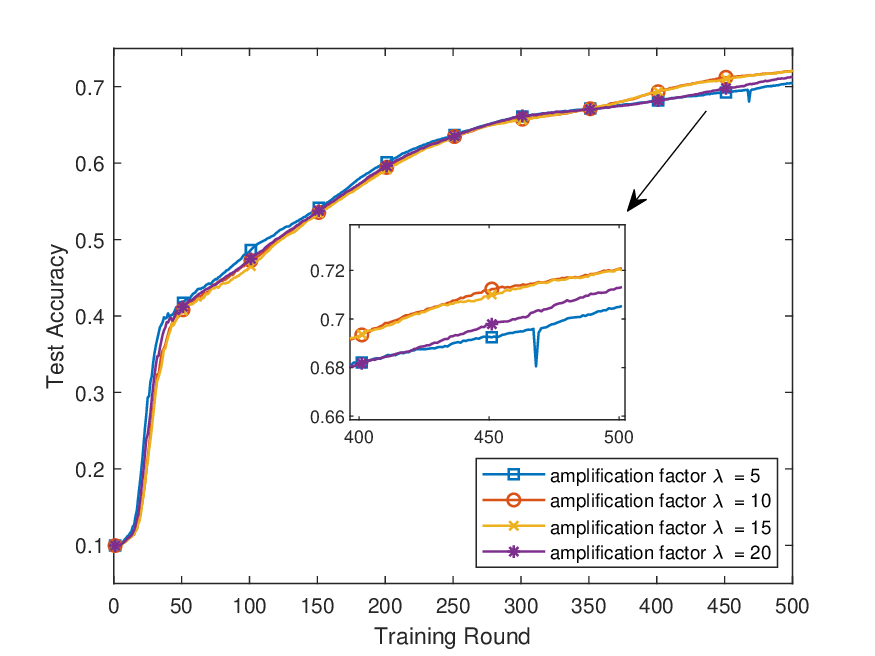}
        \caption{Text accuracy versus $\lambda$ with set 1.}
        \label{Fig. 10}
    \end{center}
\end{figure}
In Fig. \ref{Fig. 10}, we illustrate the impact of different ASTARS amplification factors on the learning rate to highlight the advantages of ASTARS-assisted FL. One observation is that the learning rate of ASTARS-assisted FL reaches its optimal value at amplification factors of 10 and 15. This is attributed to the amplification effect of ASTARS, which enhances the signal, allowing for better transmission to the BS. However, when the amplification factor is set to 5, the optimal result cannot be achieved under Simulation Condition 1, as signal loss occurs during transmission. On the other hand, when the amplification factor is too high, thermal noise becomes the dominant factor, resulting in a decline in learning accuracy. This suggests that optimizing the choice of amplification factor could be a promising direction for future research.

\section{Conclusion}\label{Conclusion} 
This paper has investigated the feasibility of the ASTARS uplink network utilizing OTA computing and has examined the learning performance of FL within this network. Specifically, we have derived an upper bound on the performance of FL by analyzing the training loss induced by communication noise. Subsequently, we have applied the SCA algorithm to jointly optimize the ASTARS phase shift and the received beam assignment. Simulation results have highlighted the advantages of the proposed ASTARS-FL scheme. In particular, the ASTARS-SCA-FL has achieved a test accuracy exceeding 72\%, closely approaching the performance of the ideal Noise-free-FL. Compared to PSTARS-FL and RIS-FL, ASTARS-based methods have attained approximately 58\% and 50\% accuracy, respectively. Furthermore, the proposed ASTARS-SCA-FL has outperformed other ASTARS variants, such as ASTARS-MM-FL and ASTARS-DC-FL, by approximately 3\% and 5\%, respectively, underscoring the effectiveness of the joint optimization strategy in mitigating performance degradation caused by communication noise.
In addition, the learning performance of the ASTARS-assisted FL scheme has remained superior across different numbers of ASTARS elements, further validating its robustness and scalability.

\appendices

\section*{Appendix~A: Proof of Theorem \ref{theorem:2}} \label{Appendix:A}
\renewcommand{\theequation}{A.\arabic{equation}}
\setcounter{equation}{0}
Since $F\left( {{{\bf{w}}_t}} \right)$ satisfies assumptions \eqref{F strongly convex}-\eqref{twice-continuously differentiable}, the following inequality relation exists
\begin{align}\label{simplify c1}
F\left( {{{\bf{w}}_{t + 1}}} \right) \leqslant F\left( {{{\bf{w}}_t}} \right) + \frac{L}{2}\left\| {{{\bf{w}}_{t + 1}} - {{\bf{w}}_t}} \right\|_2^2 \nonumber \\ 
+ {\left( {{{\bf{w}}_{t + 1}} - {{\bf{w}}_t}} \right)^T}\nabla F\left( {{{\bf{w}}_t}} \right),
\end{align}
Substituting ${{\bf{w}}_{t + 1}} = {{\bf{w}}_t} - \frac{1}{L}{{\bf{r}}_t}$ into \eqref{simplify c1} and simplifying gives
\begin{align}\label{simplify c2}
F\left( \mathbf{w}_{t+1} \right)&\leqslant F\left( \mathbf{w}_t \right) + \left( \mathbf{w}_t - \frac{1}{L}\mathbf{r}_t - \mathbf{w}_t \right)^T \nabla F\left( \mathbf{w}_t \right) \nonumber\\
&+ \frac{L}{2} \left\| \mathbf{w}_t - \frac{1}{L}\mathbf{r}_t - \mathbf{w}_t \right\|_2^2 \\ \nonumber
&= F\left( \mathbf{w}_t \right) - \frac{1}{L} \left( \mathbf{r}_t \right)^T \nabla F\left( \mathbf{w}_t \right) + \frac{1}{2L} \left\| \mathbf{r}_t \right\|_2^2. 
\end{align}

The true gradient is equal to the sum of the gradient and the error computed by  BS, i.e., ${{\bf{r}}_t} = \nabla F\left( {{{\bf{w}}_t}} \right) + {{\bf{e}}_t}$. Substituting this into \eqref{simplify c2}, we can obtain
\begin{align}\label{simplify c3}
F\left( {{{\mathbf{w}}_{t + 1}}} \right) &\leqslant F\left( {{{\mathbf{w}}_t}} \right) - \frac{1}{L}{\left( {\nabla F\left( {{{\mathbf{w}}_t}} \right) + {{\mathbf{e}}_t}} \right)^T}\nabla F\left( {{{\mathbf{w}}_t}} \right) \nonumber \\
&+ \frac{1}{{2L}}\left\| {\nabla F\left( {{{\mathbf{w}}_t}} \right) + {{\mathbf{e}}_t}} \right\|_2^2 \nonumber \\
&= F\left( {{{\mathbf{w}}_t}} \right) - \frac{1}{L}\left\| {\nabla F\left( {{{\mathbf{w}}_t}} \right)} \right\|_2^2 - \frac{1}{L}\nabla F{\left( {{{\mathbf{w}}_t}} \right)^T}{{\mathbf{e}}_t} + \nonumber\\
&\frac{1}{{2L}}\left\| {\nabla F\left( {{{\mathbf{w}}_t}} \right)} \right\|_2^2 + \frac{1}{L}\nabla F{\left( {{{\mathbf{w}}_t}} \right)^T}{{\mathbf{e}}_t} + \frac{1}{{2L}}\left\| {{{\mathbf{e}}_t}} \right\|_2^2 \nonumber\\
&= F\left( {{{\mathbf{w}}_t}} \right) - \frac{1}{{2L}}\left\| {\nabla F\left( {{{\mathbf{w}}_t}} \right)} \right\|_2^2 + \frac{1}{{2L}}\left\| {{{\mathbf{e}}_t}} \right\|_2^2.
\end{align}
The expectation processing of the communication noise and a simple mathematical treatment gives \eqref{wt+1 upper limit}. The proof is complete.

\section*{Appendix~B: Proof of Lemma \ref{theorem:3}} \label{Appendix:B}
\renewcommand{\theequation}{B.\arabic{equation}}
\setcounter{equation}{0}
The communication noise causes the error $\mathbb{E}\left[ {\left\| {{{\mathbf{e}}_{2,t}}} \right\|_2^2} \right]$, so its expression is \eqref{equ:BigWrited LiZi}.
\begin{figure*}[t]
{\small
\begin{align} \label{equ:BigWrited LiZi}
\mathbb{E}\left[ {\left\| {{{\mathbf{e}}_{2,t}}} \right\|} \right] = \frac{1}{{{{\left( {\sum\nolimits_{\varphi  \in \Phi } {{K_\varphi }} } \right)}^2}}}\sum\limits_{\varphi  = 1}^\Phi  {\mathbb{E}\left\{ {\left[ {\frac{{\left( {{{\left| {{K_\varphi } - {{\mathbf{f}}^H}\left( {{{\mathbf{h}}_n}{s_{n\Delta }} + {{\mathbf{h}}_n}{s_{m\Delta }}} \right)} \right|}^2} + \left( {{{\left| {{{\mathbf{f}}^H}{{\mathbf{h}}_n}{{\mathbf{\Theta }}^n}{{\mathbf{n}}_s}} \right|}^2} + {{\left| {{{\mathbf{f}}^H}{{\mathbf{h}}_m}{{\mathbf{\Theta }}^m}{{\mathbf{n}}_s}} \right|}^2}} \right) + {{\left| {{{\mathbf{f}}^H}{{\tilde n}_0}} \right|}^2}} \right)}}{\mu }} \right]} \right\}} \nonumber \\
\mathop  \geqslant \limits^{(a)} \frac{{\sum\limits_{\varphi  = 1}^\Phi  {\mathbb{E}\left\{ {\left( {{{\left| {{{\mathbf{f}}^H}{{\mathbf{h}}_n}{{\mathbf{\Theta }}^n}{{\mathbf{n}}_s}} \right|}^2} + {{\left| {{{\mathbf{f}}^H}{{\mathbf{h}}_m}{{\mathbf{\Theta }}^m}{{\mathbf{n}}_s}} \right|}^2}} \right) + {{\left| {{{\mathbf{f}}^H}{{\tilde n}_0}} \right|}^2}} \right\}} }}{{\mu {{\left( {\sum\nolimits_{\varphi  \in \Phi } {{K_\varphi }} } \right)}^2}}} = \frac{{\sum\limits_{\varphi  = 1}^\Phi  {\left\{ {\mathbb{E}\left( {{{\left| {{{\mathbf{h}}_{sr}}{{\mathbf{\Theta }}^n}{{\mathbf{n}}_s}} \right|}^2}} \right) + \mathbb{E}\left( {{{\left| {{{\mathbf{h}}_{sr}}{{\mathbf{\Theta }}^m}{{\mathbf{n}}_s}} \right|}^2}} \right)} \right\} + \Phi \sigma _0^2} }}{{\mu {{\left( {\sum\nolimits_{\varphi  \in \Phi } {{K_\varphi }} } \right)}^2}}}.
\end{align}
}
\hrulefill
\end{figure*}
Since ${K_\varphi }$ is the local dataset, and all the signals transmitted after reflection by ASTARS are received by the BS, ${K_\varphi }$ is equal to the user's signal after transmission through the channel. Thus, we obtain (a). Since ${\mathbf{\Theta} _n} = \sqrt {\lambda {\beta _r}} diag\left( {{e^{j\theta _1^n}}, \cdots ,{e^{j\theta _q^m}}, \cdots ,{e^{j\theta _Q^r}}} \right)$, ${\mathbf{\Theta} _m} = \sqrt {\lambda {\beta _t}} diag\left( {{e^{j\theta _1^m}}, \cdots ,{e^{j\theta _q^m}}, \cdots ,{e^{j\theta _Q^m}}} \right)$, ${{\mathbf{h}}_{sr}} = \sqrt {{\eta _0}d_{sr}^{ - \alpha }} {[h_{sr}^1, \cdots ,h_{sr}^l, \cdots ,h_{sr}^L]^H}$, $h_{sr}^q = \sqrt {\frac{\kappa }{{\kappa  + 1}}}  + \sqrt {\frac{1}{{\kappa  + 1}}} \tilde h_{sr}^q$, $\tilde h_{sr}^q \sim \mathcal{C}\mathcal{N}\left( {0,1} \right)$, and $\mathbb{E}{\left| {{\mathbf{h}}_{sr}^H{\mathbf{\Theta} _n}{{\mathbf{n}}_s}} \right|^2}$ can be rewritten as
\begin{align}\label{noise of Hn}
\mathbb{E}\left[ {{{\left| {{{\mathbf{h}}_{sr}}{{\mathbf{\Theta }}_n}{{\mathbf{n}}_s}} \right|}^2}} \right] &= \mathbb{E}\left[ {\lambda {\beta _r}{\eta _0}d_{sr}^{ - \alpha }\sigma _s^2J{{\left| {\sum\nolimits_{q = 1}^Q {h_r^q} } \right|}^2}} \right] \notag \\
 &= \lambda {\beta _r}{\eta _0}d_{sr}^{ - \alpha }JQ\left( {\frac{{Q\kappa  + 1}}{{\kappa  + 1}}} \right)\sigma _s^2.
\end{align}
Similarly, we can obtain
\begin{align}\label{noise of Hm}
\mathbb{E}{\left| {{{\bf{h}}_{sr}}{\bf{\Theta}}_m{{\bf{n}}_s}} \right|^2} = \lambda {\beta _t}{\eta _0}d_{sr}^{ - \alpha }JQ\left( {\frac{{Q\kappa  + 1}}{{\kappa  + 1}}} \right)\sigma _s^2.
\end{align}

Substituting \eqref{noise of Hn}  and \eqref{noise of Hm} into \eqref{equ:BigWrited LiZi}, we can obtain
\begin{align}\label{simplify2 noise of Hn}
\mathbb{E}\left[ {\left\| {{{\mathbf{e}}_{2,t}}} \right\|} \right] = \frac{{\Phi \left( {\tau \sigma _s^2 + \sigma _0^2} \right)}}{{\mu {{\left( {\sum\nolimits_{\varphi  \in \Phi } {{K_\varphi }} } \right)}^2}}},
\end{align}
where $\tau $ is a constant and $\tau  \triangleq \lambda \left( {{\beta _t} + {\beta _r}} \right){\eta _0}d_{sr}^{ - \alpha }JQ\left( {\frac{{Q\kappa  + 1}}{{\kappa  + 1}}} \right)$. $\mu $ is obtained through the association of \eqref{average transmit power}, \eqref{transmit sequence}, and \eqref{linear estimator}: $|{p_\varphi }{|^2} = \frac{{K_\varphi ^2\mu \nu _\varphi ^2}}{{{{\left| {{{\mathbf{f}}^H}{{\mathbf{h}}_n}} \right|}^2} + {{\left| {{{\mathbf{f}}^H}{{\mathbf{h}}_m}} \right|}^2}}} \leqslant {P_\varphi }$, and $\mu  = \mathop {\min }\limits_{_{\varphi  \in \Phi }} \frac{{{P_\varphi }{{\left| {{{\mathbf{f}}^H}{{\mathbf{h}}_n}} \right|}^2} + {{\left| {{{\mathbf{f}}^H}{{\mathbf{h}}_m}} \right|}^2}}}{{K_\varphi ^2\nu _\varphi ^2}}$.

Lemma 1 can be obtained by bringing $|{p_\varphi }{|^2}$ and $\mu$ in \eqref{simplify2 noise of Hn}, followed by some straightforward mathematical operations. The proof is
complete.
\section*{Appendix~C: Proof of Theorem \ref{theorem2}} \label{Appendix:C}
\renewcommand{\theequation}{C.\arabic{equation}}
\setcounter{equation}{0}
Based on \eqref{e2,t}, we can further analyze $K_\varphi ^2\nu _\varphi ^2$
\begin{align}\label{kv}
\nonumber
\Phi K_\varphi ^2\nu _\varphi ^2 =& K_\varphi ^2\sum\limits_{\varphi  = 1}^\Phi  {\left( {{g_\varphi }\left[ d \right] - {{\bar g}_\varphi }} \right)} \leqslant K_\varphi ^2\sum\limits_{\varphi  = 1}^\Phi  \left( {{g_\varphi }\left[ d \right]} \right) \\ \nonumber
&= \left\| {\sum\nolimits_{\left( {{{\mathbf{x}}_{\varphi k}},{y_{\varphi k}}} \right) \in {\Phi _\varphi }} {\nabla f\left( {{\mathbf{w}};{{\mathbf{x}}_k},{y_k}} \right)} } \right\| _2^2 \\ \nonumber
&\leqslant {\left( {\sum\nolimits_{\left( {{{\mathbf{x}}_{\varphi k}},{y_{\varphi k}}} \right) \in {\Phi _\varphi }} {{{\left\| {\nabla f\left( {{\mathbf{w}};{{\mathbf{x}}_k},{y_k}} \right)} \right\|}_2}} } \right)^2} \\ \nonumber
&\mathop \leqslant \limits^{(a)} {\left( {\sum\nolimits_{\left( {{{\mathbf{x}}_{\varphi k}},{y_{\varphi k}}} \right) \in {\Phi _\varphi }} {\sqrt {{\alpha _1} + {\alpha _2}\left\| {\nabla F\left( {{{\mathbf{w}}_t}} \right)} \right\|_2^2} } } \right)^2} \nonumber\\
&= K_\varphi ^2\left( {{\alpha _1} + {\alpha _2}\left\| {\nabla F\left( {{{\mathbf{w}}_t}} \right)} \right\|_2^2} \right),
\end{align}
where (a) is obtained by \eqref{upper bonuded}, $\mathbb{E}\left[ {\left\| {{{\mathbf{e}}_{2,t}}} \right\|} \right]$ can be written as
\begin{align}\label{e2 smplify}
\nonumber
\mathbb{E}\left[ {\left\| {{{\mathbf{e}}_{2,t}}} \right\|_2^2} \right] \leqslant& \frac{{\tau \sigma _s^2 + \sigma _0^2}}{{{P_\varphi }{{\left( {\sum\nolimits_{\varphi  \in \Phi } {{K_\varphi }} } \right)}^2}}}\left( {{\alpha _1} + {\alpha _2}\left\| {\nabla F\left( {{{\mathbf{w}}_t}} \right)} \right\|_2^2} \right)\\
 &\times \mathop {\max }\limits_{\varphi  \in \Phi } \frac{{K_\varphi ^2}}{{{{\left| {{{\mathbf{f}}^H}{{\mathbf{h}}_n}} \right|}^2} + {{\left| {{{\mathbf{f}}^H}{{\mathbf{h}}_m}} \right|}^2}}}.
\end{align}

By jointly solving \eqref{more simplify F(w+1)} and \eqref{e2 smplify} and performing a series of mathematical transformations, we can obtain \eqref{e[f]},
\begin{figure*}[t]
\begin{align}\label{e[f]}
\nonumber
\mathbb{E}\left[ {F({{\mathbf{w}}_{t + 1}})} \right] =& \mathbb{E}\left[ {F({{\mathbf{w}}_t})} \right] - \frac{{\left\| {\nabla F({{\mathbf{w}}_t})} \right\|_2^2}}{{2L}}\left( {1 - {\alpha _2}\frac{{\tau \sigma _s^2 + \sigma _0^2}}{{{P_\varphi }{{\left( {\sum\nolimits_{\varphi  \in \Phi } {{K_\varphi }} } \right)}^2}}}\mathop {\max }\limits_{\varphi  \in \Phi } \frac{{K_\varphi ^2}}{{{{\left| {{{\mathbf{f}}^H}{{\mathbf{h}}_n}} \right|}^2} + {{\left| {{{\mathbf{f}}^H}{{\mathbf{h}}_m}} \right|}^2}}}} \right)  \nonumber\\
&+ \frac{{{\alpha _1}}}{{2L}}\frac{{\tau \sigma _s^2 + \sigma _0^2}}{{{P_\varphi }{{\left( {\sum\nolimits_{\varphi  \in \Phi } {{K_\varphi }} } \right)}^2}}}\mathop {\max }\limits_{\varphi  \in \Phi } \frac{{K_\varphi ^2}}{{{{\left| {{{\mathbf{f}}^H}{{\mathbf{h}}_n}} \right|}^2} + {{\left| {{{\mathbf{f}}^H}{{\mathbf{h}}_m}} \right|}^2}}}.
\end{align}
\end{figure*}
where $d\left\{ {Q,{\bf{f}},{\mathbf{\Theta }^n},{\mathbf{\Theta }^m}}\right\} \triangleq \frac{{\tau \sigma _s^2 + \sigma _0^2}}{{{P_\varphi }{{\left( {\sum\nolimits_{\varphi  \in \Phi } {{K_\varphi }} } \right)}^2}}}\mathop {\max }\limits_{\varphi  \in \Phi } \frac{{K_\varphi ^2}}{{{{\left| {{{\mathbf{f}}^H}{{\mathbf{h}}_n}} \right|}^2} + {{\left| {{{\mathbf{f}}^H}{{\mathbf{h}}_m}} \right|}^2}}}$ and $\Upsilon \left\{ {Q,{\mathbf{f}},{\mathbf{\Theta }^n},{\mathbf{\Theta }^m}} \right\} \triangleq \left\{ {1 - \frac{\rho }{L} + \frac{\rho }{L}{\alpha _2}d\left\{ {Q,{\mathbf{f}},{\mathbf{\Theta }^n},{\mathbf{\Theta }^m}} \right\}} \right\}$, and then simplifying \eqref{e[f]}, we obtain \eqref{d and f}. After $t+1$ iterations, we can obtain \eqref{fw-f0}. The proof is complete.
\begin{figure*}[t]
{\small
\begin{align}\label{d and f}
\nonumber
\mathbb{E}\left[ {F({{\mathbf{w}}_{t + 1}}) - F({{\mathbf{w}}^*})} \right] \leqslant &\mathbb{E}\left[ {F({{\mathbf{w}}_t})} \right] - \mathbb{E}\left[ {F({{\mathbf{w}}^*})} \right] - \frac{{\rho \mathbb{E}\left[ {\left( {F({{\mathbf{w}}_t}) - F({{\mathbf{w}}^*})} \right)} \right]}}{L}\left( {1 - {\alpha _2}d\left\{ {Q,{\mathbf{f}},{{\mathbf{\Theta }}^n},{{\mathbf{\Theta }}^m}} \right\}} \right) + \frac{{{\alpha _1}}}{{2L}}d\left\{ {Q,{\mathbf{f}},{{\mathbf{\Theta }}^n},{{\mathbf{\Theta }}^m}} \right\} \\
&= \mathbb{E}\left[ {F({{\mathbf{w}}_t}) - F({{\mathbf{w}}^*})} \right]\left[ {1 - \frac{\rho }{L} + \frac{\rho }{L}{\alpha _2}d\left\{ {Q,{\mathbf{f}},{{\mathbf{\Theta }}^n},{{\mathbf{\Theta }}^m}} \right\}} \right] + \frac{{{\alpha _1}}}{{2L}}d\left\{ {Q,{\mathbf{f}},{{\mathbf{\Theta }}^n},{{\mathbf{\Theta }}^m}} \right\}.
\end{align}
}
\hrulefill
\end{figure*}
\section*{Appendix~D: Proof of Theorem \ref{theorem3}} \label{Appendix:D}
\renewcommand{\theequation}{D.\arabic{equation}}
\setcounter{equation}{0}
Utilising the gradient descent lemma and {\cite[Lemma~1]{bertsekas2000gradient}}, we can get
\begin{align}\label{fx+1}
  F&\left( {{{\bar \phi }_{n,m}}[k + 1]} \right) \leqslant F\left( {{{\bar \phi }_{n,m}}[k]} \right) \notag \\
  &+ \frac{{{\alpha _i}}}{I}\nabla F{\left( {{{\bar \phi }_{n,m}}[k]} \right)^T}\sum\limits_{i = 1}^I {\left( {\phi _{n,m}^{i,{\text{inx}}}[k] - {{\bar \phi }_{n,m}}[k]} \right)}  \notag \\
  &+ \frac{L}{2}{\left( {\frac{{{\alpha _i}}}{I}} \right)^2}{\sum\limits_{i = 1}^I {\left\| {\phi _{n,m}^{i,{\text{inx}}}[k] - \phi _{n,m}^i[k]} \right\|} ^2}.
\end{align}

With the aid of the auxiliary term $\frac{{{\alpha _i}}}{I}\nabla F{\left( {{{\bar \phi }_{n,m}}[k]} \right)^T}\sum\limits_{i = 1}^I {\left( {\tilde \phi _{n,m}^i[k] + \tilde \phi _{n,m}^{i,{\text{av}}}[k] + \hat \phi _{n,m}^i\left( {{{\bar \phi }_{n,m}}[k]} \right)} \right)} $, we obtain
\begin{align}\label{fx2}
  F&\left( {\bar \phi _{n,m}^{i + 1}[k + 1]} \right) \leqslant F\left( {{{\bar \phi }_{n,m}}[k]} \right) \notag \\
   &+ \frac{{\alpha [k]}}{I}\nabla F{\left( {{{\bar \phi }_{n,m}}[k]} \right)^T}\sum\limits_{i = 1}^I {\left( {\hat \phi _{n,m}^i\left( {{{\bar \phi }_{n,m}}[k]} \right) - {{\bar \phi }_{n,m}}[k]} \right)}  \notag\\
  & + \frac{{\alpha [k]}}{I}\nabla F{\left( {{{\bar \phi }_{n,m}}[k]} \right)^T}\sum\limits_{i = 1}^I {\left( {\tilde \phi _{n,m}^{i,{\text{av}}}[k] - \hat \phi _{n,m}^i\left( {{{\bar \phi }_{n,m}}[k]} \right)} \right)}  \notag \\
  & + \frac{{{\alpha _i}}}{I}\nabla F{\left( {{{\bar \phi }_{n,m}}[k]} \right)^T}\sum\limits_{i = 1}^I {\left( {\tilde \phi _{n,m}^i[k] - \tilde \phi _{n,m}^{i,{\text{av}}}[k]} \right)}  \notag \\
   &+ \frac{{{\alpha _i}}}{I}\nabla F{\left( {\bar \phi _{n,m}^i} \right)^T}\sum\limits_{i = 1}^I {\left( {\phi _{n,m}^{i,{\text{inx}}}[k] - \tilde \phi _{n,m}^i[k]} \right)}  \notag \\
   &+ \frac{L}{2}{\left( {\frac{{{\alpha _i}}}{I}} \right)^2}{\sum\limits_{i = 1}^I {\left\| {\phi _{n,m}^{i,{\text{inx}}}[k] - \phi _{n,m}^i[k]} \right\|} ^2}. 
\end{align}
The upper bound of the second term is
\begin{align}\label{upfx}
  &\frac{{\alpha [k]}}{I}\nabla F{\left( {{{\bar \phi }_{n,m}}[k]} \right)^T}\sum\limits_{i = 1}^I {\left( {\hat \phi _{n,m}^i\left( {{{\bar \phi }_{n,m}}[k]} \right) - {{\bar \phi }_{n,m}}[k]} \right)}  \notag \\
  &\mathop  \leqslant \limits^{(a)}  - \frac{{{c_\tau }}}{I}{\alpha _i}\sum\limits_{i = 1}^I {{{\left\| {\hat \phi _{n,m}^i\left( {{{\bar \phi }_{n,m}}[k]} \right) - {{\bar \phi }_{n,m}}[k]} \right\|}^2}}  \notag \\
  &+ \alpha [k]\left( {G\left( {{{\bar \phi }_{n,m}}[k]} \right) - \frac{1}{I}\sum\limits_{i = 1}^I {G\left( {\hat \phi _{n,m}^i\left( {{{\bar \phi }_{n,m}}[k]} \right)} \right)} } \right) \notag \\
  &\mathop  \leqslant \limits^{(b)}  - \frac{{{c_\tau }}}{I}{\alpha _i}\sum\limits_{i = 1}^I {{{\left\| {\hat \phi _{n,m}^i\left( {{{\bar \phi }_{n,m}}[k]} \right) - {{\bar \phi }_{n,m}}[k]} \right\|}^2}}  \notag \\
  &+ G\left( {{{\bar \phi }_{n,m}}[k]} \right) - G\left( {{{\bar \phi }_{n,m}}[k + 1]} \right) \notag \\
  & + \frac{{\alpha [k]}}{I}\sum\limits_{i = 1}^I {G\left( {\phi _{n,m}^{i,{\text{inx}}}[k]} \right) - G\left( {\hat \phi _{n,m}^i\left( {{{\bar \phi }_{n,m}}[k]} \right)} \right)}  \notag \\
  &\mathop  \leqslant \limits^{(c)}  - \frac{{{c_\tau }}}{I}{\alpha _i}\sum\limits_{i = 1}^I {{{\left\| {\hat \phi _{n,m}^i\left( {{{\bar \phi }_{n,m}}[k]} \right) - {{\bar \phi }_{n,m}}[k]} \right\|}^2}}  \notag \\
  &+ G\left( {{{\bar \phi }_{n,m}}[k]} \right) - G\left( {{{\bar \phi }_{n,m}}[k + 1]} \right) \notag \\
  & + \frac{{{L_G}}}{I}{\sum\limits_{i = 1}^I {\left\| {\phi _{n,m}^{i,{\text{inx}}}[k] - \phi _{n,m}^i[k]} \right\|} ^2},
\end{align}
where $G$ is a convex function (possibly not differentiable) with bounded subgradient function, and there exists ${L_G} > 0$ such that $\left\| {\partial G\left( x \right)} \right\| \leqslant {L_G}$.
The right-hand side of (a) is obtained from {\cite[Proposition~5]{facchinei2015parallel}} ${\left( {{{{\rm{\hat x}}}_i}\left( z \right) - z} \right)^T}\nabla F\left( z \right) + G\left( {{{{\rm{\hat x}}}_i}\left( z \right)} \right) - G\left( z \right) \le  - {c_\tau }{\left\| {{{{\rm{\hat x}}}_i}\left( z \right) - z} \right\|^2}$, where ${c_\tau } \buildrel \Delta \over = {\min _i}{\tau _i} > 0$; in (b) is obtained from $G\left( {{{\bar \phi }_{n,m}}[k + 1]} \right) \le \left( {1 - \alpha [k]} \right)G\left( {{{\bar \phi }_{n,m}}[k + 1]} \right) + \frac{{\alpha [k]}}{I}\sum\limits_{i = 1}^I {G\left( {\phi _{n,m}^{i,{\rm{inx}}}[k]} \right)} $; and (c) is based on $\left\| {\partial G\left( {\phi _{n,m}^{i,{\rm{inx}}}[k]} \right)} \right\| \le {L_G}$.
By substituting \eqref{upfx} into \eqref{fx2}, and applying the trigonometric inequality along with the bound $\left\| {{\rm{x}}_i^{{\rm{inx}}}[n] - {{\rm{x}}_i}[n]} \right\| \le c$, where $c > 0$, we can write
\begin{align}\label{Ux}
U&\left( {{{\bar \phi }_{n,m}}[k + 1]} \right)\leqslant U\left( {{{\bar \phi }_{n,m}}[k]} \right) + \frac{{{c^2}{L_{\max }}}}{{2I}}{\alpha ^2}\left[ k \right]\notag \\
&- \frac{{{c_\tau }}}{I}\alpha \left[ k \right]{\sum\limits_{i = 1}^I {\left\| {\hat \phi _{n,m}^i\left( {{{\bar \phi }_{n,m}}[k]} \right) - {{\bar \phi }_{n,m}}[k]} \right\|} ^2} \notag \\
&+ {c_1}\alpha \left[ k \right]\sum\limits_{i = 1}^I {\left\| {\tilde \phi _{n,m}^{i,{\text{av}}}[k] - \hat \phi _{n,m}^i\left( {{{\bar \phi }_{n,m}}[k]} \right)} \right\|} \notag \\
&+ {c_1}\alpha \left[ k \right]\sum\limits_{i = 1}^I {{\varepsilon _i}\left[ n \right]}  + {c_1}\alpha \left[ k \right]\sum\limits_{i = 1}^I {\left\| {\tilde \phi _{n,m}^i[k] - \tilde \phi _{n,m}^{i,{\text{av}}}[k]} \right\|}, 
\end{align}
where ${c_1} = {L_G}/I$. We now apply {\cite[Lemma~1]{bertsekas2000gradient}} with the following identifications: $Y[k] = U\left( {{{\bar \phi }_{n,m}}[k]} \right)$, $X[k] = \frac{{{c_\tau }}}{I}\alpha \left[ k \right]{\sum\limits_{i = 1}^I {\left\| {\hat \phi _{n,m}^i\left( {{{\bar \phi }_{n,m}}[k]} \right) - {{\bar \phi }_{n,m}}[k]} \right\|} ^2}$ and $Z[k] = {c_{1}}\alpha \left[ k \right]\sum\limits_{i = 1}^I {\left\| {\tilde \phi _{n,m}^{i,{\text{av}}}[k] - \hat \phi _{n,m}^i\left( {{{\bar \phi }_{n,m}}[k]} \right)} \right\|} 
 + {c_{1}}\alpha \left[ k \right]\sum\limits_{i = 1}^I {{\varepsilon _i}\left[ n \right]}  + {c_{1}}\alpha \left[ k \right]\sum\limits_{i = 1}^I {\left\| {\tilde \phi _{n,m}^i[k] - \tilde \phi _{n,m}^{i,{\text{av}}}[k]} \right\|}  + \frac{{{c^2}{L_{\max }}}}{{2I}}{\alpha ^2}\left[ k \right]$.
Since $\sum\nolimits_{k = 1}^\infty Z[k] < \infty$ and ${\lim _{\left\| {\operatorname{x}  \to \infty } \right\|}}U\left( \operatorname{x}  \right) =  + \infty $, it follows that $U\left( \bar{\phi}_{n,m}[k] \right)$ converges to a finite value and $\sum\nolimits_{n = 1}^\infty  {\alpha [k]\hat \phi _{n,m}^i\left( {{{\bar \phi }_{n,m}}[k]} \right) - {{\bar \phi }_{n,m}}[k] < \infty } ,\forall i = 1, \cdots ,I$. Using \eqref{eqtheorem3}, we can obtain $\mathop {\lim \inf }\limits_{k \to \infty } \left\| {\hat \phi _{n,m}^i\left( {{{\bar \phi }_{n,m}}[k]} \right) - {{\bar \phi }_{n,m}}[k]} \right\| = 0,\forall i = 1, \cdots ,I$.

Based on \cite{facchinei2015parallel}, it is easy to show that $\mathop {\lim \sup }\limits_{k \to \infty } \left\| {\hat \phi _{n,m}^i\left( {{{\bar \phi }_{n,m}}[k]} \right) - {{\bar \phi }_{n,m}}[k]} \right\| = 0,\forall i = 1, \cdots ,I$. Hence, we have $\mathop {\lim }\limits_{k \to \infty } \left\| {\hat \phi _{n,m}^i\left( {{{\bar \phi }_{n,m}}[k]} \right) - {{\bar \phi }_{n,m}}[k]} \right\| = 0,\forall i = 1, \cdots ,I$. Since the sequence ${\left\{ {{{\bar \phi }_{n,m}}[k]} \right\}_k}$ is bounded, there exists a limit point within the feasible domain.
In order for $\mathop {\lim }\limits_{k \to \infty } \left\| {\hat \phi _{n,m}^i\left( {{{\bar \phi }_{n,m}}[k]} \right) - {{\bar \phi }_{n,m}}[k]} \right\| = 0,\forall i = 1, \cdots ,I$ to hold, $\hat \phi _{n,m}^i\left( {\bar \phi _{n,m}^\infty [k]} \right) = \bar \phi _{n,m}^\infty [k]$ must be satisfied, and $\bar \phi _{n,m}^\infty [k]$ is also a steady-state solution to the optimization problem.
This completes the proof of both parts \textit{a)}  and \textit{b)}  of the theorem.

\bibliographystyle{IEEEtran}
\bibliography{mybib}

\end{document}